\documentclass[fleqn,usenatbib,useAMS]{mnras}

\usepackage{graphicx}	% Including figure files
\usepackage{amsmath}	% Advanced maths commands
\usepackage{amssymb}	% Extra maths symbols
\usepackage{multicol}        % Multi-column entries in tables
\usepackage{bm}		% Bold maths symbols, including upright Greek
\usepackage{pdflscape}	% Landscape pages

\newcommand{\lsun}{\ifmmode{{\rm ~L}_\odot}\else{~L$_\odot$}\fi}
\newcommand{\Msun}{~M$_\odot$}
\newcommand{\ujybm}{\,$\mu$Jy/beam}
\newcommand{\ujy}{\,$\mu$Jy}

\usepackage[T1]{fontenc}
\usepackage{ae,aecompl}

\usepackage{newtxtext,newtxmath}
\title[Odd Radio Circles]{MeerKAT uncovers the physics of an Odd Radio Circle}
\author[Ray P. Norris, et al.]
{Ray P. Norris$^{1,2}$
\thanks{Contact e-mail: \href{mailto:raypnorris@gmail.com}{raypnorris@gmail.com}},
J. D. Collier$^{1,3,10}$,
Roland M. Crocker $^4$,
Ian Heywood$^5$,
\newauthor
Peter Macgregor$^{1,2}$,
L. Rudnick$^{6}$,
Stas Shabala$^{7}$,
Heinz Andernach$^8$,
\newauthor
Elisabete da Cunha$^{11}$,
Jayanne English$^9$,
Miroslav Filipovi\'c$^1$,
B\"arbel S. Koribalski$^{1,2}$,
\newauthor
Kieran Luken$^{1,13}$,
Aaron Robotham$^{11,12}$,
Srikrishna Sekhar$^3$,
Jessica E. Thorne$^{11}$,
\newauthor
Tessa Vernstrom$^{10}$
\\
$^{1}$School of Science, Western Sydney University, Locked Bag 1797, Penrith, NSW 2751, Australia\\
$^{2}$CSIRO Space \& Astronomy, Australia Telescope National Facility,  P.O. Box 76, Epping, NSW 1710, Australia\\
$^{3}$The Inter-University Institute for Data Intensive Astronomy (IDIA), Department of Astronomy, University of Cape Town, Private Bag X3, Rondebosch, 7701, South Africa\\
$^{4}$Research School of Astronomy and Astrophysics, Australian National University, Canberra 2611, A.C.T., Australia\\
$^{5}$Sub-Dept. of Astrophysics, Department of Physics, University of Oxford, Denys Wilkinson Building, Keble Rd., Oxford, OX1 3RH, UK\\
$^6$ Minnesota Institute for Astrophysics, University of Minnesota, 116 Church St. SE, Minneapolis, MN 55455, USA\\
$^{7}$ School of Natural Sciences, University of Tasmania, Private Bag 37, Hobart, TAS 7001, Australia\\
$^8$ Depto.\ de Astronom\'{i}a, DCNE, Universidad de Guanajuato, Cj\'on.\ de Jalisco s/n, Guanajuato, CP 36023, Mexico\\
$^{9}$ Department of Physics and Astronomy, University of Manitoba, Winnipeg, MB R3T 2N2, Canada\\
$^{10}$CSIRO Space \& Astronomy,  Australia Telescope National Facility, PO Box 1130, Bentley WA 6102, Australia\\
$^{11}$International Centre for Radio Astronomy Research, M468, University of Western Australia, Crawley, WA 6009, Australia\\
$^{12}$ARC Centre of Excellence for Astrophysics in Three Dimensions (ASTRO3D)\\
$^{13}$CSIRO Data61, P.O. Box 76, Epping, NSW 1710, Australia\\
}
\date{Last updated xxxx; in original form xxxx}

\pubyear{2021}

\begin{document}
\label{firstpage}
\pagerange{\pageref{firstpage}--\pageref{lastpage}}
\maketitle

\begin{abstract}
Odd Radio Circles (ORCs) are recently-discovered faint diffuse circles of radio emission, of unknown cause, surrounding galaxies at moderate redshift ($z\sim0.2-0.6$). 
Here we present detailed new MeerKAT radio images at 1284 MHz of the first ORC, originally discovered with the Australian Square Kilometre Array Pathfinder, with higher resolution (6 arcsec) and sensitivity ($\sim$ 2.4 \ujybm).
 In addition to the new images, which reveal a complex internal structure consisting of multiple arcs, we also present polarisation and spectral index maps.  Based on these new data, we consider potential mechanisms that may generate the ORCs. 
\end{abstract}

%\clearpage

%%%%%%%%%%%%%%%%%%%%%%%%%%%%%%%%%%%%%%%%%%%%%%%%%%

%%%%%%%%%%%%%%%%% BODY OF PAPER %%%%%%%%%%%%%%%%%%

% The MNRAS class isn't designed to include a table of contents, but for this document one is useful.
% I therefore have to do some kludging to make it work without masses of blank space.
% \begingroup
% \let\clearpage\relax
% \setcounter{tocdepth}{3}
% \tableofcontents
% \endgroup
% \newpage
% \clearpage

\section{Introduction}

Odd Radio Circles (ORCs) appear as circles of diffuse radio emission typically about one arcmin in diameter. So far, no corresponding diffuse emission at non-radio wavelengths (optical, IR, UV, X-ray) has been detected. In some cases the ORCs have a  galaxy at a redshift  $z \sim 0.3-0.6$ at their centre, which, if associated (see  Section \ref{comparison}),  implies an ORC diameter of hundreds of kpc.

ORCs were first discovered \citep{norris20} during the Pilot Survey \citep{norris21} of the Evolutionary Map of the Universe \citep[EMU:][]{emu}, using the Australian Square Kilometre Array Pathfinder (ASKAP) telescope \citep{hotan21}. A further ORC was  discovered by Intema \citep{norris20} in data from the Giant Metrewave Radio Telescope (GMRT), and an additional ORC has since been discovered, also in ASKAP data, by \citet{koribalski21}. 

The ORCs strongly resemble supernova remnants or planetary nebulae, but their Galactic latitude distribution rules out a Galactic origin \citep{norris20}. They also resemble the rings of radio emission sometimes seen in nearby starburst galaxies \citep[e.g. NGC6935;][]{norris20}, but the absence of an optical counterpart to the diffuse radio emission rules out this explanation. \citet{norris20} discuss and reject a number of other potential causes, favouring a spherical shell from a transient event.
\citet{koribalski21} discuss three of these hypotheses for the origin of ORCs:
\begin{itemize}
    \item a spherical shock wave from the central galaxy, perhaps from a cataclysmic event such as a merger of SMBH (super-massive black hole), or else the termination shock of a starburst wind,
    \item a double-lobed radio Active Galactic Nucleus (AGN) seen end-on,
    \item the result of interactions between galaxies.
\end{itemize}
Other suggested explanations for ORCS include the throats of wormholes \citep{kirillov20, kirillov21}.
However, at present, there is no agreed explanation for the mechanism that produces them.

\begin{figure*}
%\begin{center}
\includegraphics[width=18cm, angle=0]{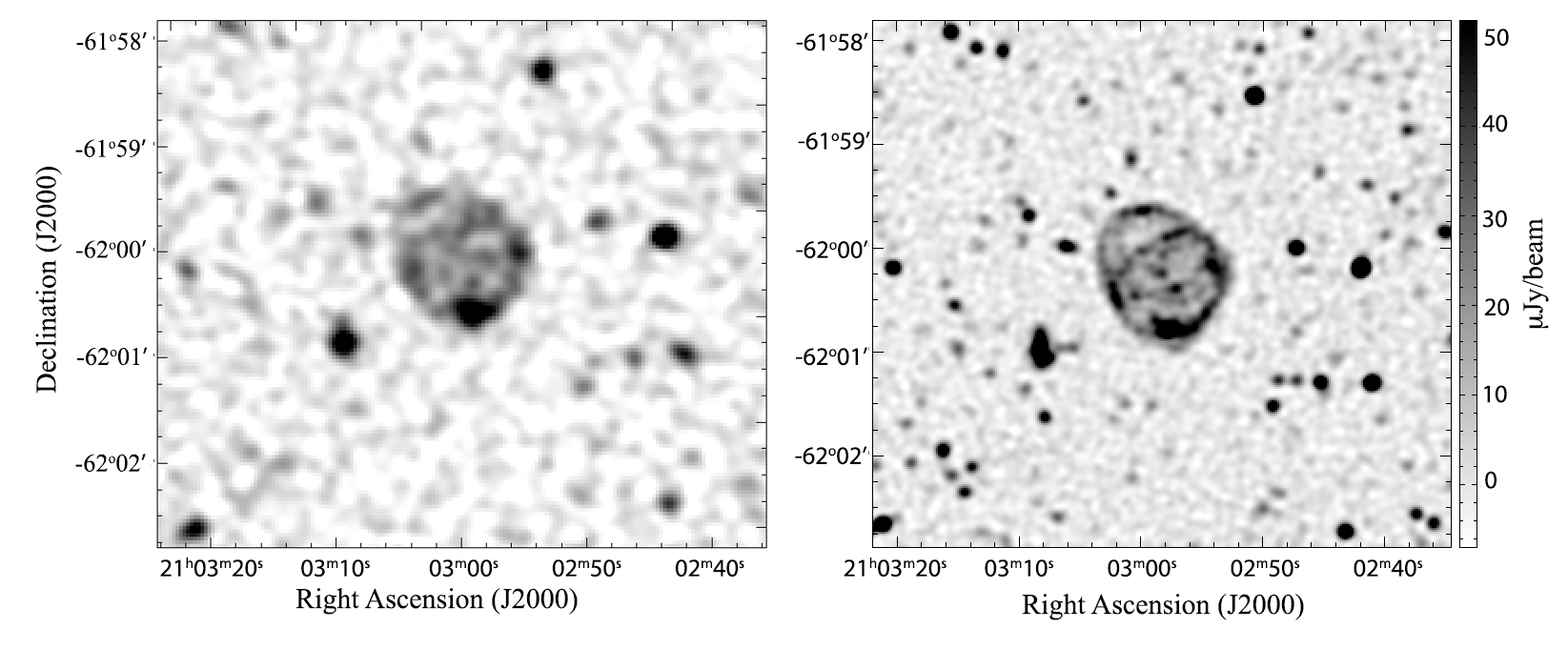}
\caption{(Left) The ASKAP image of ORC1,
adapted from \citet{norris20}  at 944 MHz. The resolution is 11 arcsec, and the rms sensitivity is 25 \ujybm. (Right) The new Meerkat image of ORC1  at 1284 MHz. The resolution is 6 arcsec, and the rms sensitivity is 2.4 \ujybm.  A FITS file of these images is available in the online supplementary material.
 }
\label{askap1}
%\end{center}
\end{figure*}

Here we describe  1284 MHz observations with the MeerKAT  telescope of the first ORC (ORC J2103-6200, also known as ORC1) to be recognised, in an attempt to test these and other hypotheses.
ORC1 was discovered  at 944 MHz using the ASKAP telescope, but was subsequently  observed \citep{norris20} using the Australia Telescope Compact Array \citep[ATCA:][]{wilson11}  at 2.1 GHz and the Murchison Widefield Array \citep[MWA:][]{tingay13} at  88--154 MHz, establishing its reality beyond doubt.
The ASKAP image is shown in Figure \ref{askap1}.

\begin{table*}
\centering
\caption{Properties of the optical/IR sources associated with the ORCs.  }
\setlength{\tabcolsep}{2pt} % reduces space between columns to fit on the page horizontally
\renewcommand{\arraystretch}{1.25} % increases space between rows to make it more readable
\label{tab:optical}
\begin{tabular}{lcccccccccccccccccl}
\hline
 & & radio &Observing & Fractional & 1.4 GHz & &
 %\multicolumn{2}{c}{GALEX} &
 \multicolumn{4}{c}{DES/SDSS} & \multicolumn{3}{c}{WISE} & \\
Source Name  & ID  & flux density & Frequency & radio & luminosity & SFR &
% FUV & NUV & 
g & r & i & z &  W1 & W2 & W3 &  $z_{phot}$ \\
 & & [mJy] & (MHz) & luminosity  &W/Hz & \Msun/yr &
% \multicolumn{2}{c}{[mag]} &
\multicolumn{4}{c}{[AB mag]} & 
\multicolumn{3}{c}{[Vega mag]} & \\
\hline
WISEA J210258.15--620014.4 & ORC1C &  0.091 & 1284& 0.027& 7.3$\times10^{22}$  & 40.2
& 22.04 & 20.10 & 19.23 & 18.79 &  15.07  & 14.98  & $>$12.94  & 0.55  \\
% & & & & & & & 0.06 & 0.01 & 0.01 & 0.02 & 0.04 & $\pm$0.031 & $\pm$0.061 & & & \\
WISEA J210257.88--620046.3 & ORC1S & 0.74 & 1284& 0.222& 5.9$\times 10^{23}$ & 326.6 & 19.73 & 18.95 & 18.55 & 18.35 & 15.47  & 15.06  & 11.20  &  0.23 \\
WISEA J155524.65+272633.7 & ORC4C & 1.43 & 325 & 0.026 &3.7 $\times 10^{23}$ & 203.5& 21.18 & 19.64 & 19.00 & 18.40 & 15.50 & 15.31  & >13.19  &  0.46 \\
WISEA J010224.35--245039.6 & ORC5C & 0.1 & 944 & 0.050 &1.6$\times 10^{22}$ & 8.8
& 20.43 & 18.97 & 18.48 & 18.18 & 15.39 & 15.19 & $>$12.36 & 0.27 \\
\hline
\end{tabular}
\raggedright{ORC1C, ORC4C, ORC5C are the central galaxies in their respective ORCs. ORC1S is the galaxy superimposed on the south of the ring in ORC1, which we consider to be unlikely to be physically associated with the ORC because of the very different redshift. Fractional luminosity is the flux density of the galaxy divided by the flux density of the diffuse emission of the ORC. Radio luminosity is calculated at 1.4 GHz to enable comparison with other results, and to calculate SFR (the calculated star formation rate if all the radio emission of the galaxy were due to star formation) using the relation derived by \citet{bell03}.} Redshifts are from \citet{zou20}.
\end{table*}

Sections 2 and 3 of this paper describe the observations and data reduction respectively, the results of which are presented in Section 4. In Section 5, we discuss the phenomenology of ORCs, and in Section 6 we discuss the nature and environment of the host galaxy. Guided by these new results, in Section 7 we discuss possible interpretations of the ORC phenomenon.

\section{Observations}

The MeerKAT radio telescope \citep{jonas16, mauch20} is the South African precursor for the Square Kilometre Array. It consists of 64 13.5–m diameter antennas with single pixel, offset Gregorian feeds. 70\% of the antennas are contained within a 1–km inner core, while the remaining antennas are distributed over a wider area giving a maximum baseline length of 8 km. The observations described here have  a centre frequency of 1284 MHz and a bandwidth of 800 MHz.

We observed ORC1 with MeerKAT on 6 June 2020 as a Director's Discretionary Time observation (proposal number DDT-20200519-RN-01). The standard calibrator PKS B1934--638 was observed for 5 minutes for every 30 minute target scan, and there was a single 10 minute scan of the primary polarisation calibrator 3C 286 (J1331+3030). The total on-source time for ORC1 was 10 hours. 

\section{Data Reduction}
 As the  data processing techniques for Meerkat data are still being optimised, we tried two approaches in parallel. The first was known to produce excellent Stokes I images but  did not do polarisation or spectral calibration, whereas the second approach calibrated for polarisation and spectral index. In the second approach, the data is imaged separately in a number of different spectral windows, resulting in a slightly lower overall sensitivity than the first approach.

In both cases, the first step is to convert the visibilities from their native format to Measurement Set format, averaging down to 1024 frequency channels to reduce the data volume. 

We then flagged (excised) those frequency channels that are affected by radio frequency interference,   including the removal of known persistent radio frequency interference regions, and then autoflagged data using  using the the {\sc casa flagdata} task of {\sc casa} package \citep{mcmullin2007}.

About 50\% of the data were flagged because of  RFI (Radio Frequency Interference).  Most of the losses occurred on short spacings in the geolocation satellite bands at about 1150--1300 and 1520--1610 MHz, with secondary losses at the lower end of the band around 930 MHz due to aviation and GSM. More detail on the RFI environment at Meerkat can be found in \citet{mauch20}.

In both approaches, we used PKS B1934--638 as the bandpass, flux and phase calibrator,  using the flux scale measured by \citet{reynolds94}, and, in the second approach, used J1331+3030 (3C286) as the polarisation calibrator. 

\subsection{Approach 1 (Imaging)}
In the first approach, we derived flux, delay, bandpass and time-dependent gain corrections from calibrator observations via an iterative process with rounds of flagging based on the residual visibilities between iterations.  We applied the calibration solutions  to the target visibilities, which were then flagged using the {\sc tricolour}\footnote{\url{https://github.com/ska-sa/tricolour/}} package. 

We imaged the data twice using {\sc wsclean} \citep{wsclean1}, with the second round of imaging using a deconvolution mask based on the results of the first round. We derived phase and delay self-calibration solutions using {\sc cubical} \citep{kenyon2018}. We then imaged the self-calibrated visibilities using {\sc ddfacet} \citep{tasse2018}, with direction dependent corrections derived using {\sc killms} \cite[e.g.][]{smirnov2015}. No primary beam corrections were applied, since the target of interest is at the phase centre and has a small angular extent (1 arcmin) compared to the FWHM (full-width at half-maximum) of 66 arcmin of the primary beam. The final image was produced with a \citet{briggs1995} weighting of $-$0.3, resulting in a rms noise of 2.4 $\mu$Jy beam$^{-1}$ at the image centre, and an angular resolution  (FWHM of the restoring beam) of 6$''$.2~$\times$~5$''$.9, PA $-$88 deg. The processing scripts are available online if further details are required \citep{heywood2020}.

This processing resulted in an image of ORC1 which is about ten times deeper than  the EMU Pilot Survey image, with twice the resolution.

\subsection{Approach 2: Spectral Index and Polarisation}
\label{sec:spindexpol}
 The second approach followed a similar overall strategy to the first, but included  polarisation and spectral index calibration, and resulted in spectro-polarimetry cubes. We processed the data with the \textsc{CASA}-based \citep{mcmullin2007} IDIA pipeline (Collier et al., in prep.)\footnote{\url{https://idia-pipelines.github.io/docs/processMeerKAT}} at the
\emph{ilifu}\footnote{\url{http://www.ilifu.ac.za/}} research facility.

 The first step was to split the data into 11 spectral windows (SPWs), shown in Table \ref{tab:SPW}, whose edges were chosen to  exclude the known frequency ranges of persistent RFI. We processed each SPW independently, following a standard two-round cross-calibration process. Splitting into SPWs enables a solution which is a function of frequency, accounting for the Stokes-I spectral index and any instrumental rotation in the polarisation spectrum of the calibrators. 
 
 \begin{table}
\centering
\caption{Spectral Windows (SPWs) used for spectral index and polarisation analysis. Gaps between the SPWs indicate the locations of severe RFI }
\label{tab:SPW}
\begin{tabular}{lc}
\hline
SPW \# & band \\
 & (MHz) \\
\hline
1 & 880--933  \\
2 & 960--1010 \\
3 & 1010--1060 \\
4 & 1060--1110 \\
5 & 1110--1163 \\
6 & 1299--1350 \\
7 & 1350--1400 \\
8 & 1400--1450 \\
9 & 1450--1500 \\
10 & 1500--1524 \\
11 & 1630--1680 \\
\hline
\end{tabular}
\end{table}

 We then derived flux, delay, bandpass and time-dependent gain corrections of each of the SPWs from calibrator observations via an iterative process
  This enabled better statistics for a second round of flagging, which used the same modes as before, with tighter thresholds, in addition to the `rflag` mode. 
  
  We then performed full Stokes cross-hand calibration, during which a new set of delay, bandpass, and parallel-hand gains were computed using \texttt{gaintype=`T'}, to solve for a polarisation-independent gain. Additionally, we solved for instrumental leakages using the flux calibrator, and derived the XY-phases using the polarisation calibrator.

To self-calibrate the data, we first imaged and deconvolved each SPW down to a clean limit of  100 $\mu$Jy/beam. Cleaning each SPW separately   resulted in a slight loss of signal/noise ratio compared with the first approach, but resulted in a solution which accounted  for the spectral behaviour over the field.

We then performed source finding using PyBDSF \citep{pybdsf}, from which the resulting island boundaries were used as a clean mask within a subsequent round of deconvolution down to a 10$\sigma$ threshold. We then used the output image to produce a phase-only selfcal solution for time-dependent  gains using a 1 minute solution interval and \texttt{gaintype=`T'}, which were applied and used to flag on the residual data. No primary beam corrections were applied, since the target of interest is at the phase centre and has a small angular extent (1 arcmin) compared to the beamwidth (66 arcmin FWHM).

Finally, we produced a Stokes IQUV cube for each SPW, imaging with the same input mask and threshold, and a robustness of -0.5, and a 10 arcsec Gaussian $(u,v)$ taper, in order to match the (u,v) components between each SPW, chosen to match the synthesised beam from the lowest part of the band. This was particularly important for producing the spectral index map, and averaged polarisation cube. Similarly, the images were restored with a (circular) restoring beam of 10 arcsec. 
The resulting averaged Stokes I image has an RMS of $\sim$3.5 $\mu$Jy/beam at the centre of the image. %and $\sim$2.5 $\mu$Jy/beam at the edge of the image. 

During the calibration, the polarisation of two  (SPWs 7 and 10) of the eleven SPWs were found to be slightly affected by RFI, and these were excluded from the polarisation analysis, but included in the spectral index analysis.

A similar set of cubes was produced for the calibrators, 
although over $\sim$120 channels, each 2.5 MHz wide, to verify the data quality, by comparing the IQUV spectra  to the known values for the two calibrator fields, which were found to closely match the expected values.

 A spectral index map for the source was produced by performing a least-squares fitting routine on every pixel  in the 11 SPWs within a sub-image surrounding ORC1.

\section{Results}
\subsection{Stokes-I Image}
\begin{figure*}
%\begin{center}
\includegraphics[width=20cm, angle=0]{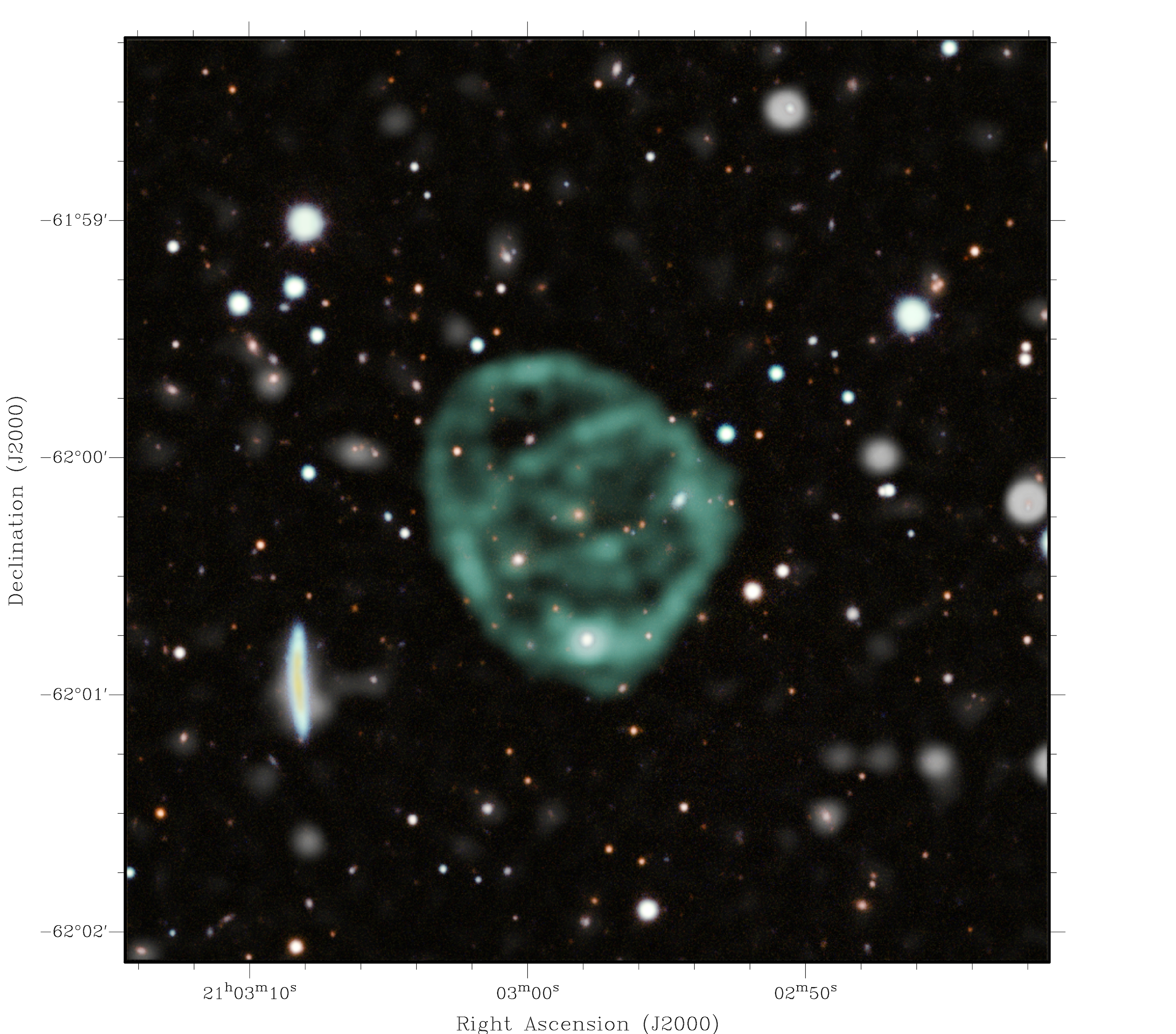}
\caption{The MeerKAT Stokes-I image of ORC1, superimposed on  optical data from the Dark Energy Survey DR1\citep{abbott18},  both spanning the same field of view. A square root transfer function was applied to the radio data of the field of view and this greyscale image was then adjusted for contrast. An image of the radio data confined to  the ORC region was assigned mint green.  This region was blended with the greyscale  radio continuum field of view image and the DES optical image  of the same field, so that
faint radio sources outside the ORC appear as faint grey diffuse patches, often surrounding their host galaxies. The filters used in the DES image were assigned turquoise, magenta, yellow and red, with the result that DES sources mainly appear in this image as white.  The layering schema employed is described in \citet{english17}. This figure is optimised to convey the structure of the ORC, and quantitative information should be  taken from Figure 1 or from the FITS files in the Supplementary Information.} 
 
\label{meerkat-des}
%\end{center}
\end{figure*}

The resulting image from the first processing stream is shown in Figures \ref{askap1} and \ref{meerkat-des}. It confirms the previous ASKAP image in showing a circular edge-brightened ring with diffuse emission inside, but in addition
\begin{itemize}
    \item it has an rms sensitivity (2.4 \ujybm) about one tenth that of the ASKAP image,
    \item the angular resolution (6 arcsec) is about twice as high as that of the ASKAP image,
    \item The central galaxy,  designated as ORC1C by \citet{norris20}, is clearly detected as a radio source. We consider this likely to be the host galaxy of the ORC, and discuss this further in Section \ref{galaxy}.
    \item The hints of internal structure seen in the ASKAP image are clearly resolved into a number of arcs of emission. 
    \item There are a number of knots of radio emission within the ORC, some of which correspond to galaxies. Some of these galaxies appear to be associated with the ORC, as discussed in Section \ref{environment} below, while others (such as the bright source ORC1S in the south of the ORC) are at very different redshifts and are presumably unassociated chance coincidences.
\end{itemize}

\subsection{Spectral Index}
Throughout this paper we define spectral index $\alpha$  in terms of the relationship between flux density $S$ and observing  frequency $\nu$  as $S \propto \nu^{\alpha}$.
 
\begin{figure}
%\begin{center}
\includegraphics[width=9cm, angle=0]{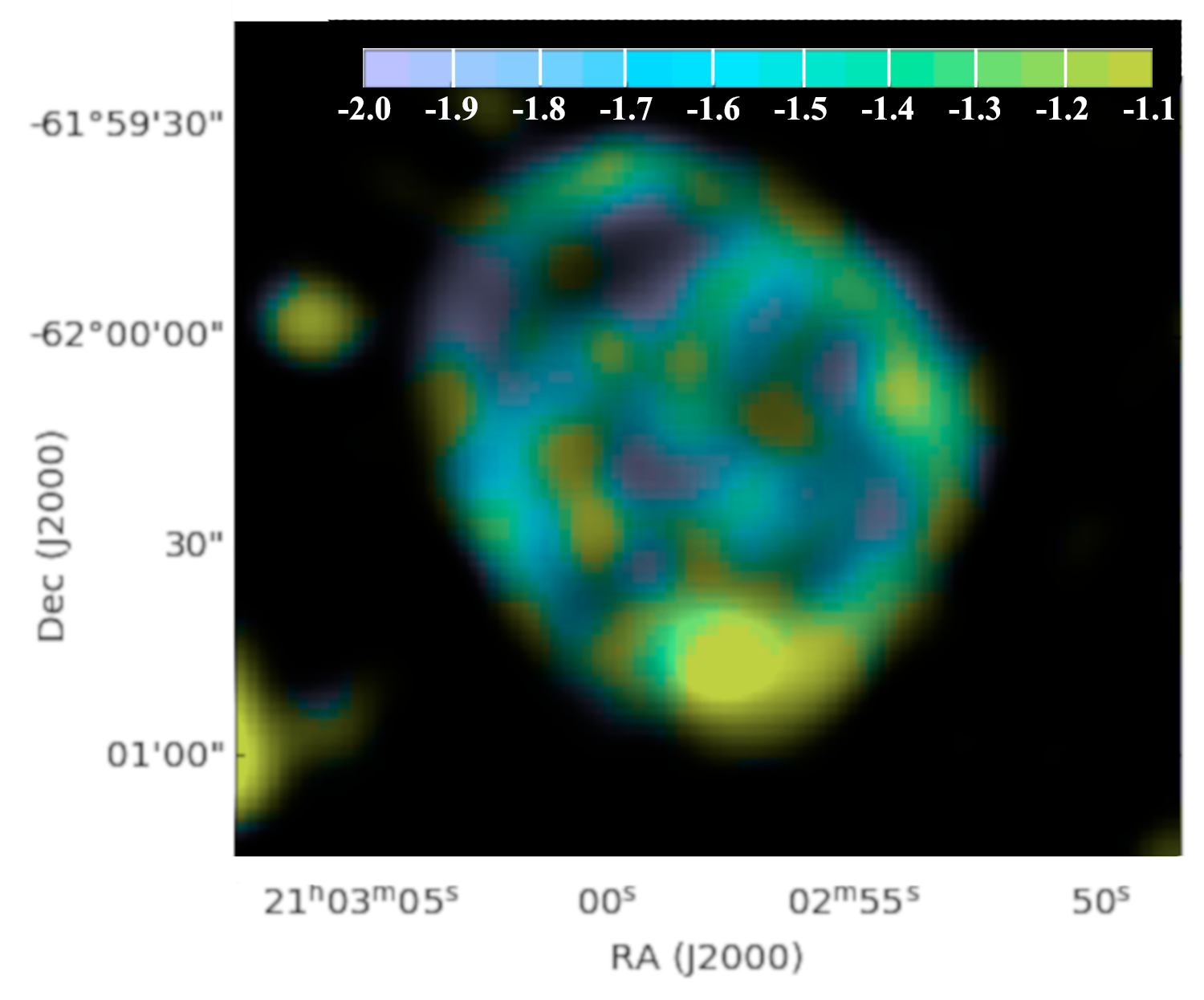}
\caption{The Meerkat spectral index image of ORC\,1. An equiluminant colour table \citep{richardson21}
%(\url{https://github.com/mlarichardson/CosmosCanvas}), 
shown in the colour bar, spans the spectral index range of -2.0 < $\alpha$ < -1.1.
Outside the  spatial region of the ORC, the purple  also represents $\alpha$ <  -2.0 while the green-yellow represents $\alpha$ > -1.1. The spectral index map has been multiplied by a square root black and white stretch of the total intensity map, which was subsequently adjusted for contrast.  We estimate the uncertainty in  spectral index to be  $\sim$0.1.
 }
\label{spindex}
%\end{center}
\end{figure}

The spectral index map,  obtained as described in Section \ref{sec:spindexpol}, is shown in Figure \ref{spindex}, which displays the spectral index using an equiluminant colour table \citep{richardson21}.
Systematics in the flux calibration in the individual frequency bands limit the accuracy of the spectral indices to $\sim \pm 0.1$.  

The ring has a median spectral index of $\sim$-1.6  but this varies considerably around the ring, as shown in Figure \ref{spindex}.  The faint regions inside the ring appear steeper, at $\sim$-1.9.  The arc structures inside the ring have a similar spectral index to the outer ring,  suggesting that they are both produced by the same mechanism.

The galaxy in the south of the ring has a spectral index of $\sim$ -0.95, and the regions of the ring adjacent to it  also have a flatter spectral index than the ring, suggesting that the radio emission from this galaxy extends over a significantly larger area than is obvious from the total intensity image. The radio luminosity of this galaxy (see Table \ref{tab:optical}) suggests that it is an AGN.

The integrated in-band spectral index of ORC1 from this data is -1.64.  \citet{norris20}  also measured this integrated spectral index using data from several telescopes. We have re-measured those integrated flux densities, and added the new flux density measurement from the Meerkat data. A least-squares fit to these data gives a multi-telescope spectral index of the integrated emission  (between 88 and 1284 MHz)  of -1.4$\pm$0.05. The discrepancy between this spectral index (-1.4)  and the in-band Meerkat spectral index (-1.64), may be attributable to spectral steepening at higher frequencies, suggesting an ageing population of electrons. For the purposes of the modelling later in this paper, we adopt the multi-telescope spectral index of -1.4.

\subsection{Polarisation}

\begin{figure}
%\begin{center}
\includegraphics[width=9cm, angle=0]{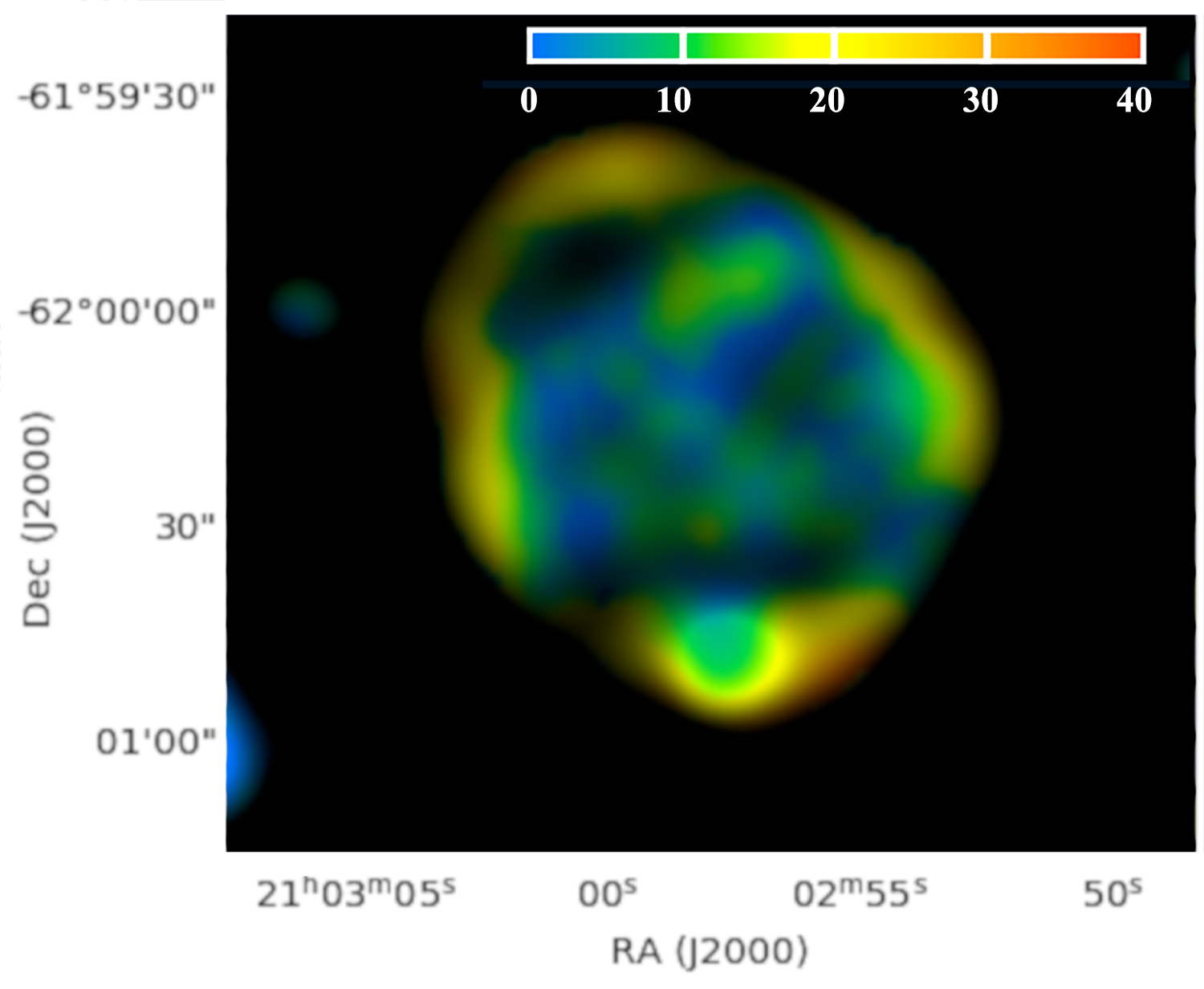}
\caption{The measured fractional polarisation of ORC\,1. The colourbar shows percentage polarisation.  To down-weight regions of low signal/noise ratio in the polarisation image, we  multiplied (i.e. masked) this colour image by the total greyscale intensity image. We estimate the uncertainty in the fractional polarisation as  $\sim$5\%.
 }
\label{poln}
%\end{center}
\end{figure}

The fractional polarisation of ORC\,1 is shown in Figure \ref{poln}. The fractional polarisation along the outer ring is as high as $\sim$30\%, with  lower fractional polarisations in the interior.

To determine the polarisation properties of the outer polarized ring, we averaged Q and U in 20$\degr$ azimuthal sectors, separately for each frequency channel, over the radial band 25 -- 50 arcsec \footnote{This radial range using a central position of  21:02:57.27, -62:00:14.3, was adopted for convenience to account for the polarised ring's slight elliptical shape.}.  We then performed a least squares fit to the resulting values of polarisation angle as a function of wavelength squared, to derive a rotation measure (RM) and electric vector polarisation angle at zero frequency for each sector.  The same procedure was done for the polarized patch in the interior.

Ignoring the unreliable values at azimuths 250$\degr$ and 350$\degr$ due to low polarized signals, the rotation measures averaged 27.2 rad/m$^2$, with an rms scatter of 4 rad/m$^2$. The Galactic foreground ranges from 17-25 rad/m$^2$ in the 3$\degr$ region surrounding ORC1 \citep{hutschenreuter21},  suggesting that there is very little contribution local to the ORC, and therefore a very dilute surrounding thermal plasma.    

\begin{figure}
%\begin{center}
\includegraphics[width=9cm, angle=0]{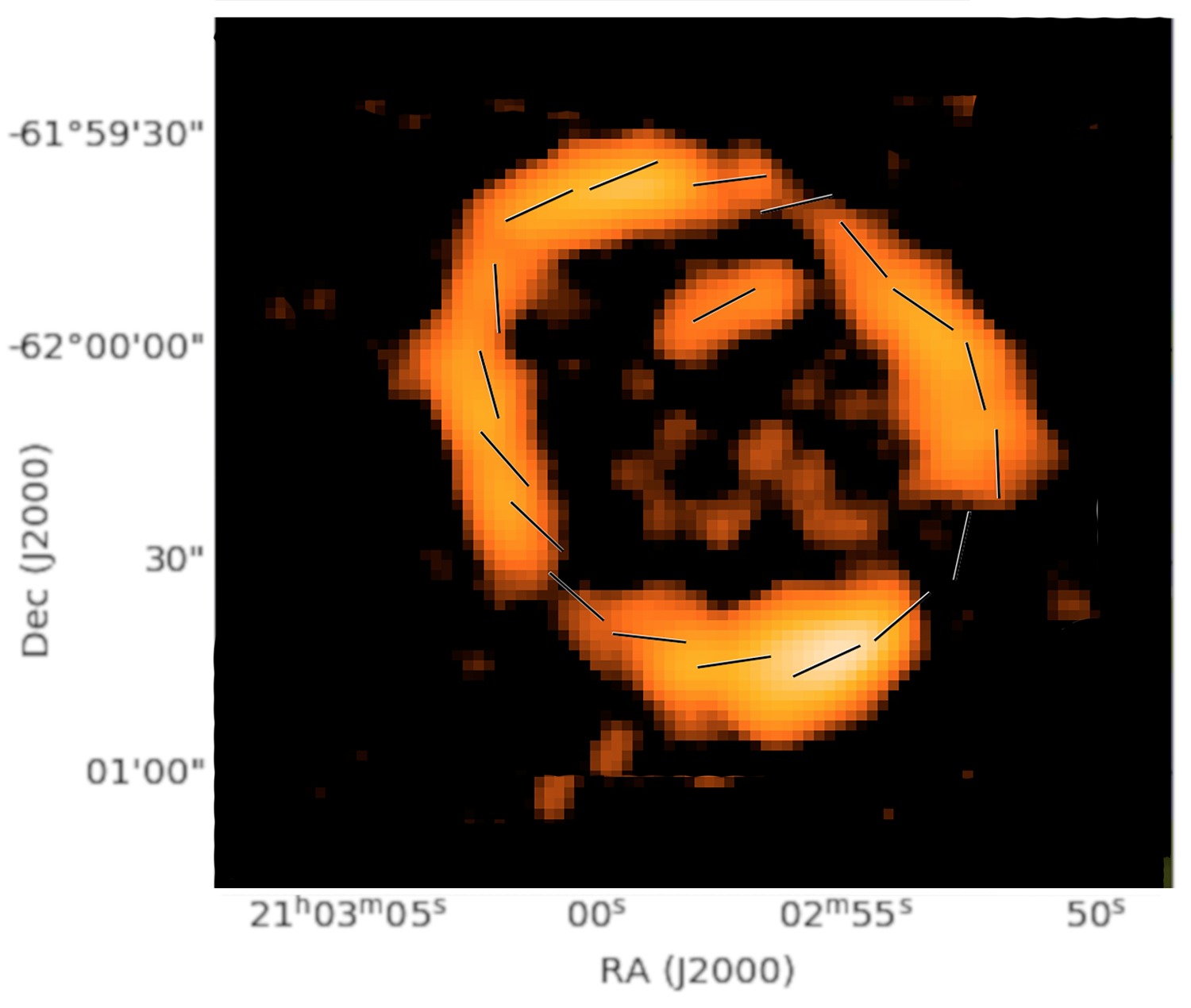}
\caption{The B vectors of the polarisation of ORC\,1, corrected to zero wavelength using the derived rotation measure, overlaid on the total polarisation image, indicating a tangential magnetic field in the ring. 
 }
\label{bvectors}
%\end{center}
\end{figure}

The resulting magnetic field vectors  (obtained by rotating the observed electric field vectors by 90\degr) are shown in Figure \ref{bvectors}.  They have a typical error of $\sim5\degr$, except for the two low polarisation azimuths, where the errors are $\sim10\degr$.  The field directions are very close to tangential to the ring structure;  an average offset of $\sim10\degr$ may be the result of uncertainties in the polarisation angle calibration.   It is also significant that the vectors vary smoothly along the ring, except for one low polarisation region. 
 
The magnetic field for the interior polarized patch is also shown;  it has a rotation measure consistent with those of the ring, and shows that the magnetic field is aligned with the direction of the arc.

We also estimated the depolarisation using the northwest half of the ring, with the caveat that the structures were somewhat different in the different frequency channels.  The values for 3 bins of 3 frequency channels each, omitting the two most unreliable channels, yielded percentage polarisations of  $\sim$27\% (for $\lambda^2$=395 cm$^2$), $\sim$21\% (658 cm$^2$), and $\sim$16\% (954 cm$^2$), indicating mild depolarisation over the observed wavelength range.  The equivalent ``k" value in a Burn depolarisation law P($\lambda$)=P(0)e$^{(-k\lambda^4)}$, which characterizes the gradients in RM,  is $\sim$ 70~rad$^2$ m$^{-4}$, similar to that seen from the medium around tailed radio galaxies \citep[e.g.][]{guidetti10}. 

\begin{figure}
%\begin{center}
\includegraphics[width=9cm, angle=0]{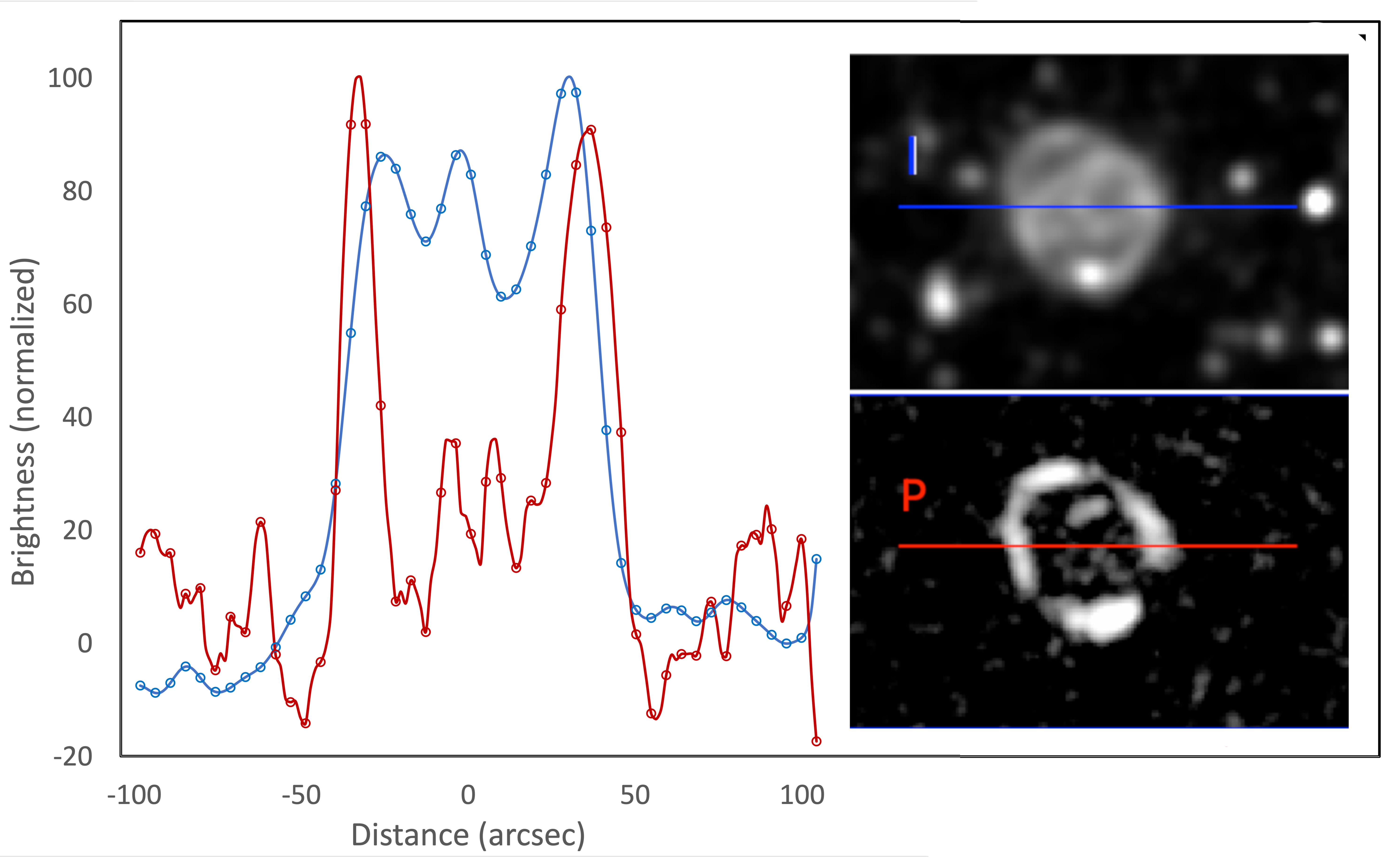}
\caption{Plots of total intensity (blue) and polarised intensity (red) along a line going through the central galaxy, as shown in the inset.  The brightness levels are the average of the first 9 frequency channels at 10 arcsec resolution, so are an approximation, and do not account for spectral index.
The plots are each normalized to their peak value in the profile, with a value of 100 representing 193 \ujybm\ for I, and 28 \ujybm\ for P.  The fractional polarisation is dramatically lower in the interior, and the peaks on the rings are at larger radii than in total intensity, leading to higher fractional polarisations on the outer edge of the ring. Data points are separated by 4.5 arcsec.
 }
\label{pslice}
%\end{center}
\end{figure}

A more detailed view of the polarisation is shown in Figure \ref{pslice}, which shows the polarisation along a cross-section through the image. This figure confirms that the  fractional polarisation in the rings is higher than that in  the interior, and also shows that the polarisation in the ring peaks  at a larger radius than in total intensity, leading to higher fractional polarisation on the outer edge of the ring.

The arc-like structures inside the ring are not as strongly polarised as the outer ring.
While this might be caused by depolarisation of an arc that is behind the ORC, arcs in the foreground would not be subject to this depolarisation. Therefore, it is likely that the lower polarisation is intrinsic to the interior arcs. 
This implies that they are physically different from the outer ring, and makes models such as shown in Figure \ref{interpretations}(a), with an equatorial and polar ring,  more attractive than models in which the structure consists of two ellipses (e.g., Figure \ref{interpretations}(b)).

\section{The Host Galaxy}
\label{galaxy}

\subsection{Environment}
\label{environment}

\citet{norris21c} have examined the environment of the three single ORCs, and found that while ORC1 is located in a significant overdensity, ORCs 4 and 5 are not, although each has a neighbour with which it is probably interacting. Here we examine the environment of ORC1 in detail.

\begin{figure*}
%\begin{center}
\includegraphics[width=17cm, angle=0]{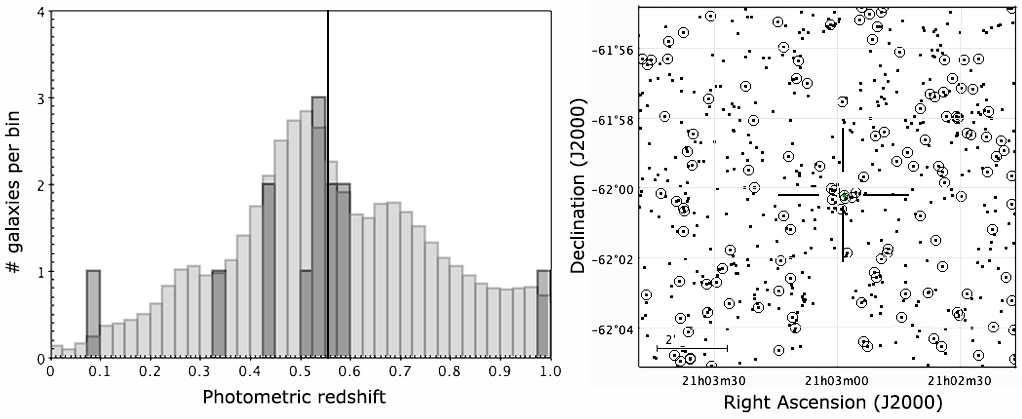}
\caption{(Left) A histogram of the photometric redshifts \citep{zou20} of the galaxies surrounding the host galaxy of ORC1, ORC1C. Dark columns shows galaxies within 0.5 arcmin of ORC1C while light columns show the (scaled) distribution of redshifts over the entire EMU Pilot Survey field. The vertical line shows the redshift of ORC1C. (Right) The galaxies surrounding ORC1. Dots show all galaxies with a photometric redshift in \citet{zou20}, and circles show those with a redshift in the range 0.5 < z < 0.6. The cross-hairs show the location of ORC1C.}
\label{cluster}
%\end{center}
\end{figure*}

In Figure \ref{cluster} we show the distribution of galaxies surrounding ORC1. It is clearly located in an overdensity of galaxies. There are 8 galaxies with 0.5 < z < 0.6  within 0.5 arcmin of ORC1, which gives a density 9 times higher than galaxies in the same redshift range in the surrounding area.
%\Ray{FYI: 56  arcmin$^2$ contains 63 galaxies.}
The individual photometric redshifts do not have enough precision (median standard error = 0.076) to accurately locate them in the radial direction (i.e. along the line of sight) relative to the ORC. However, if we  assume that the distribution in the radial direction is similar to the distribution in the tangential direction (i.e. perpendicular to the line of sight), then several of these galaxies are actually located within the shell of the ORC.

\begin{figure}
%\begin{center}
\includegraphics[width=8cm, angle=0]{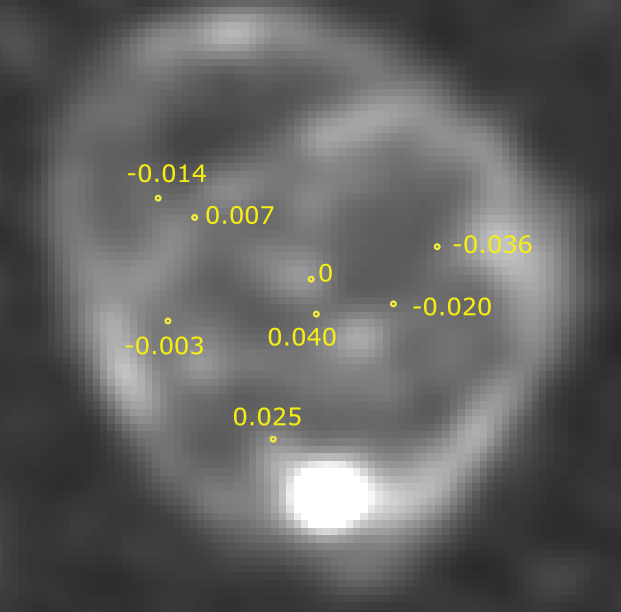}
\caption{Galaxies with  positions and photometric redshifts within the ORC, superimposed on a greyscale representation of the Meerkat continuum image. Each small circle represents one of the eight galaxies in the central overdensity shown in Figure \ref{cluster}. The number next to each is the difference between their redshift and that of ORC1C, all of which agree within 1 standard error. }
\label{orc1-galaxies}
%\end{center}
\end{figure}

In Figure \ref{orc1-galaxies} we show the distribution of the galaxies  with 0.5 < z < 0.6   within the ORC. All their redshifts agree to within 1 standard error, suggesting they are physically associated. It also suggests that several, if not all, of them are physically located within the ORC shell. We might therefore expect to see some sign of their interaction with the wind that produced the shell. No such radio structures are visible in Figure \ref{orc1-galaxies}. However, the galaxies will have moved significantly. For example, assuming a peculiar velocity of 100 km/s, they would have moved by 0.1 Mpc = 16 arcsec over the $\sim$Gyr lifetime of the ORC.  Each of them is within 16 arcsec of one of the internal arcs, so  one idea that would be interesting to explore is whether the  arcs might consist of streams of ionised gas blown off these galaxies by the starburst wind. A similar process is observed in ``Jellyfish galaxies'' \citep[e.g.][]{ebeling14}.

\subsection {The Morphology of the ORC1 host galaxy}
\begin{figure}
%\begin{center}
\includegraphics[width=8cm, angle=0]{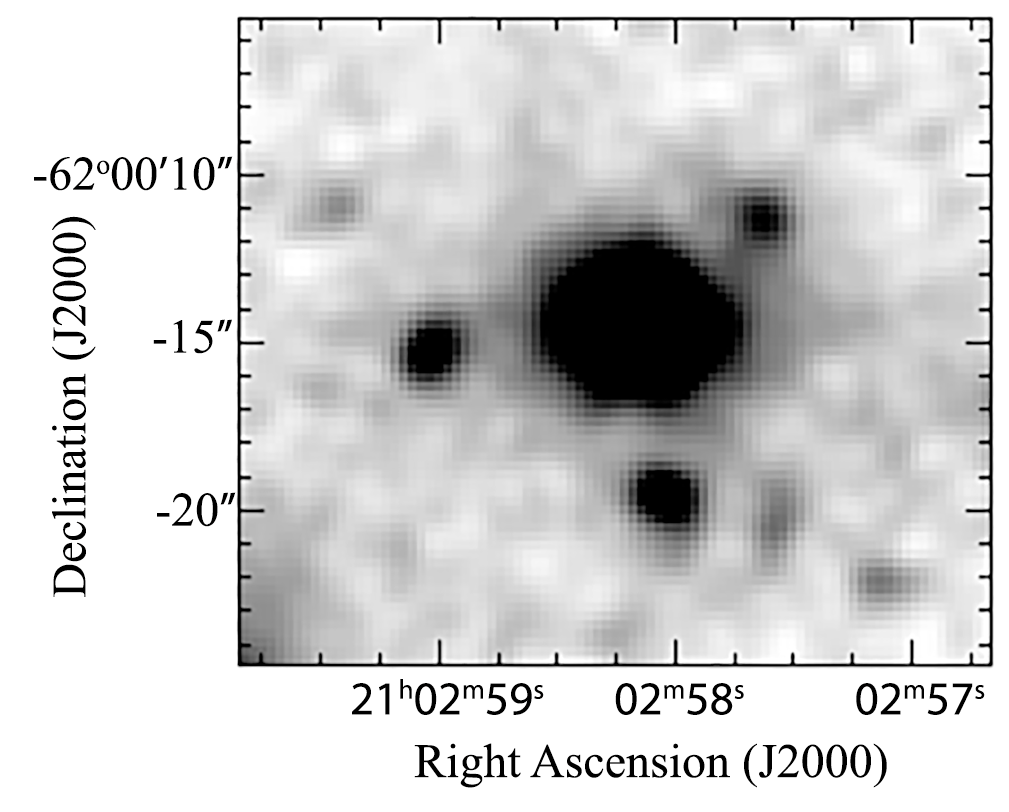}
\caption{ The  image of the ORC1 host galaxy in the I band from Dark Energy Survey data, convolved with a one arcsec Gaussian, showing an east-west extension of about 7 arcsec.
}. 
\label{residual}
%\end{center}
\end{figure}

DESI Legacy data \citep{zou20} find that the radial profile of the galaxy is best fitted by a de Vaucouleurs profile, suggesting that it is probably an elliptical galaxy.  The resulting half-light radius is 1.73 arcsec, and major and minor axes of 3.82 and 3.08 arcsec  respectively. In Figure \ref{residual} we show the  low-level emission surrounding the  central galaxy. It is evident that there is low-level diffuse emission extending to about 7 $\times$ 5 arcsec.  We speculate that this may indicate some interaction with the neighbours.

\subsection{Colours of ORC host galaxies}

\begin{figure}
%\begin{center}
\includegraphics[width=8cm, angle=0]{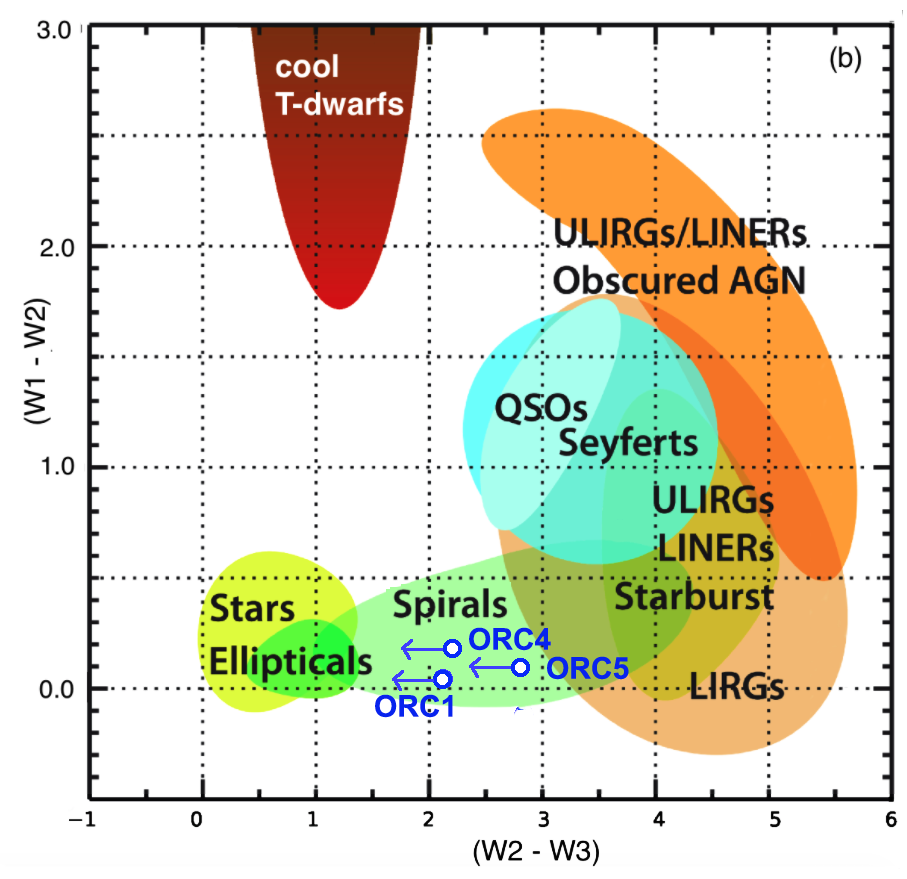}
\caption{The WISE colours of the three ORC host galaxies plotted on the WISE colour-colour diagram taken from \citet{wise} Arrows indicate upper limits.
}. 
\label{wise}
%\end{center}
\end{figure}

\begin{figure}
%\begin{center}
\includegraphics[width=8cm, angle=0]{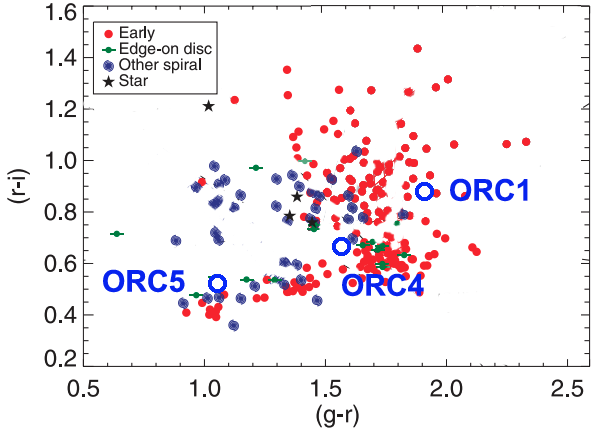}
\caption{The gri colours of the three ORC host galaxies  plotted on the gri colour-colour diagram adapted  from \citet{masters11}. gri photometry for ORC1 and ORC5 are from the Dark Energy Survey \citep{abbott18} and those for ORC4 are from SDSS \citep{sdss}
}. 
\label{optical}
%\end{center}
\end{figure}

Figure \ref{wise} show these three galaxies on a WISE colour-colour diagram. They are close together in the part of the diagram occupied by spiral galaxies, although their limits on W3 would also permit them to be ellipticals. Ideally these colours should be k-corrected. For example, if the galaxies had a Spectral Energy Distribution similar to Arp220, then the  k-correction would move ORC1 into the area occupied by LIRGs. However, we do not know the  rest-frame spectral energy distribution of these galaxies, and so these classifications must  be treated with caution.

In Figure \ref{optical} we show the gri colours of these galaxies on a diagram showing typical early (i.e. elliptical) and star-forming galaxies, adapted from \citet{masters11}. All three galaxies have colours that are redder than typical star-forming galaxies, which could either be due to dust, or to the presence of an early-type component. Again, we would ideally correct these colours for the redshift.

Table \ref{tab:optical} shows the radio luminosities of each of the galaxies, and also the star formation rate that would be inferred if the radio emission were entirely due to star formation.

\subsection{SFR of ORC1 host galaxy}
\label{sec:SFR}

Table \ref{tab:optical} shows that  the radio luminosity of the ORC1 host galaxy of $7.3\times10^{22}$ W/Hz corresponds to a star formation rate of 40 \Msun/yr, so the galaxy could either be a low-luminosity AGN or a starburst galaxy.

An alternative way of calculating the SFR is from the WISE W3 flux density, which was shown by \citet{cluver17} to provide a reasonable estimate of SFR independently of the W1 - W2 colour. WISE detected the ORC1 galaxy  only in the W1 and W2 bands, and gave a lower limit of W3 > 12.939. However, Figure \ref{wise} shows that all galaxies have W2-W3 > 0, and so we can use the measurement of W2=14.984 to place an upper limit of W3 < 14.984. We then use Equation 8 of \citet{cluver17} to derive a star formation rate in the range 4 < SFR < 20 \Msun/yr, which is rather lower than the value derived from the radio flux density (assuming the radio is not due to an AGN). 

We also used two separate SED-fitting software packages to estimate the SFR. Unfortunately there is no useful near-IR, far-IR, or ultraviolet data on this source, so the result is necessarily uncertain. In both cases, the resulting star formation rate for any reasonable fit cannot produce the observed radio flux density. Conversely, forcing the models to use the radio flux density as pure starburst resulted in a poor fit. We deduce that the radio emission is primarily due to an AGN, and we do not include it in the models. In both cases, we use the redshift as a prior.

\begin{figure}
%\begin{center}
\includegraphics[width=9cm, angle=0]{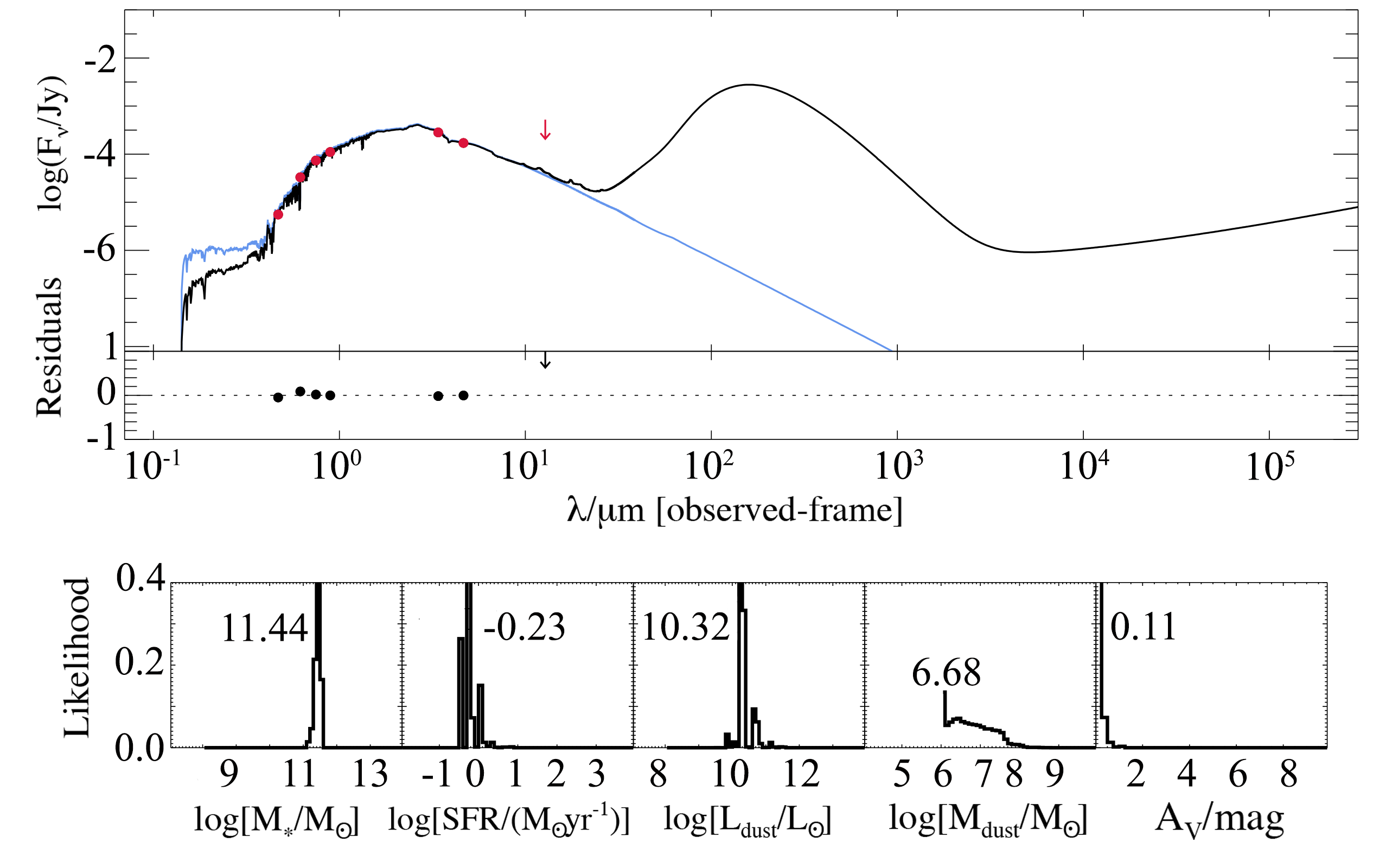}
\caption{(Upper) MAGPHYS fit to the SED of ORC1C. The blue line indicates stars, and the black line is the total. (Lower) the probability distribution from MAGPHYS of the stellar mass, the current SFR, the dust luminosity and mass, and the extinction. 
}
\label{magphys}
%\end{center}
\end{figure}

First, we used the MAGPHYS \citep{dacunha12} package, which models star-forming galaxies 
to fit the SED, resulting in the fit shown in Figure \ref{magphys}. 

\begin{figure}
%\begin{center}
\includegraphics[width=9cm, angle=0]{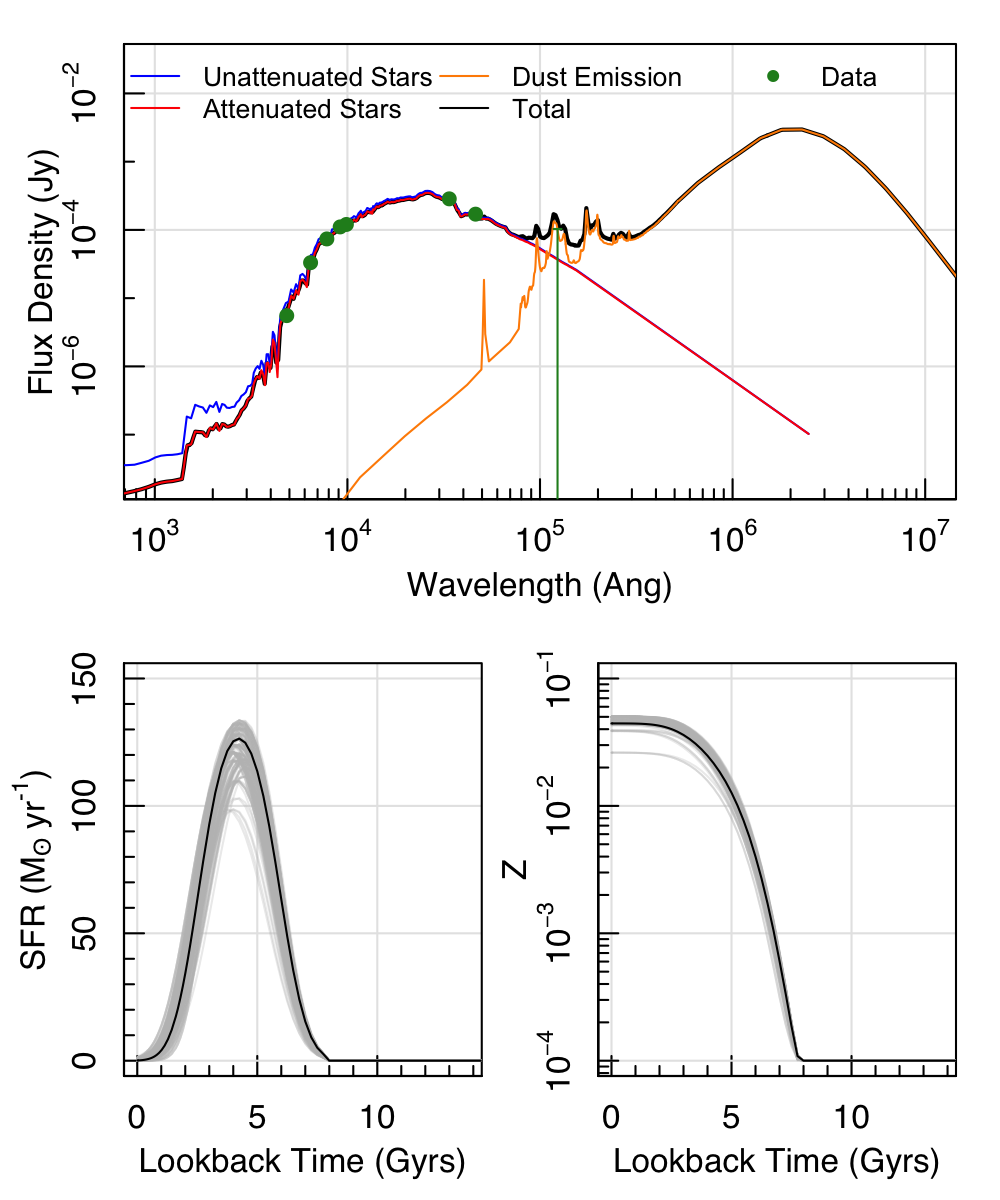}
\caption{(Upper) {\sc ProSpect} fit to the SED of ORC1C, using the photometric redshift. The green points show the input photometry and the coloured lines show the unattenuated (blue) and attenuated (red) starlight, dust emission (orange), and total galaxy SED (black). 
(Lower) The star-formation and metallicity history of the galaxy, showing a starburst about 5 Gyr ago. The grey lines show the sampling of the posterior while the black line shows the highest likelihood solution.}
\label{prospect}
%\end{center}
\end{figure}

Second, we used the {\sc ProSpect} \citep{robotham20} package and adopt a skewed normal star formation history and evolving metallicity, where the chemical enrichment is linearly mapped to the mass growth of the galaxy.
The resulting fit is shown in Figure~\ref{prospect}. 

Both codes show broad agreement 
that there is little current star formation, but evidence for a starburst several Gyr ago. The radio luminosity, and possibly the mid-IR excess, comes from an AGN. A  caveat to both these fits is that the WISE and optical photometry use different apertures, and the dust extinction is also poorly modelled without far-infrared data. The difference in the  fits between these two leading packages is an indication that our data do not provide a strong constraint on the models.

\subsection{Summary of ORC1 host properties}

There is strong circumstantial evidence (from the optical morphology, the colours, and from the SED fits) that the host galaxy is probably an elliptical galaxy, with very little current star formation, but with a strong starburst within the last few Gyr. The radio luminosity, together with the central compact source seen in the optical, then suggest that a radio-loud AGN lies at the centre of this galaxy. The ORC lies in an overdensity of galaxies with several nearby companions that lie within the sphere of the ORC, and  in future work we will explore whether these may be responsible for the arcs of radio continuum emission seen within the ORC.

\section{Phenomenology}
\subsection{Morphology and Polarisation}
\label{morphology}

\begin{figure*}
%\begin{center}
\includegraphics[width=18cm, angle=0]{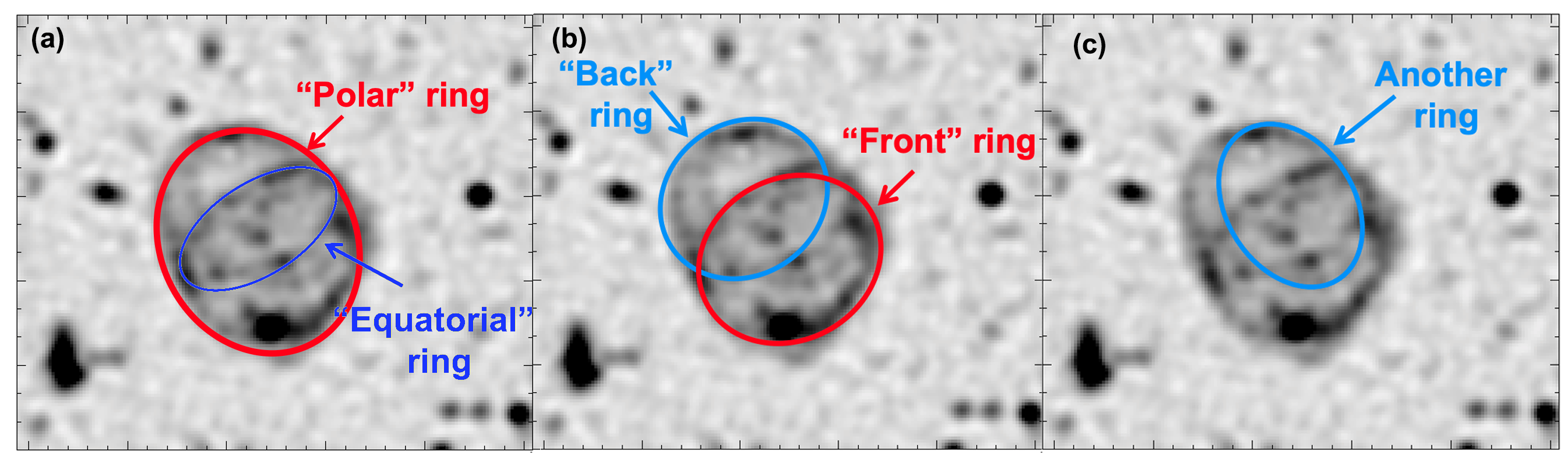}
\caption{Some potential interpretations of the structures seen within the ring of the MeerKAT image of ORC1. (a) suggests that the outer or ``polar'' ring is the edge-brightened limb of a shell, which also has an equatorial ring formed by another mechanism. (b) interprets the morphology as a front ring and a back ring, possibly caused by a biconical outflow. Finally, we note the existence of other structures within the ORC, such as a potential faint ring shown in (c), distinct from the other rings.}. 
\label{interpretations}
%\end{center}
\end{figure*}

Figure \ref{meerkat-des} shows a roughly circular outer ring, which also contains diffuse emission (i.e. the level inside the ring is slightly higher than outside the ring). The most natural explanation of this morphology is that it represents a spherical shell, discussed below. However, within the ring can be seen a number of other structures, some potential interpretations of which are shown in Figure \ref{interpretations}. A simple spherical shell model does not explain these other arcs without some modification.

 Our polarisation measurements show a high degree of polarisation along the ring, with a lower level of polarisation in the interior of the ring. The  magnetic field vectors,  shown in Figure \ref{bvectors}, show that the magnetic field directions are very close to tangential to the ring structure, and the vectors vary smoothly along the ring, except for one low polarisation region. 

The tangential magnetic field vectors are consistent with expectations for amplification by a radially-expanding shock in a  single spherical shell.

In the next few subsections we consider alternative interpretations of this morphology, and will discuss the possible mechanisms driving them in the next section.

\subsection{Spherical Shell}
We consider the scenario in which the ORC is a roughly spherical shell of optically-thin radio emission,   which we would observe as an edge-brightened ring. 

\begin{figure*}
%\begin{center}
\includegraphics[width=18cm, angle=0]{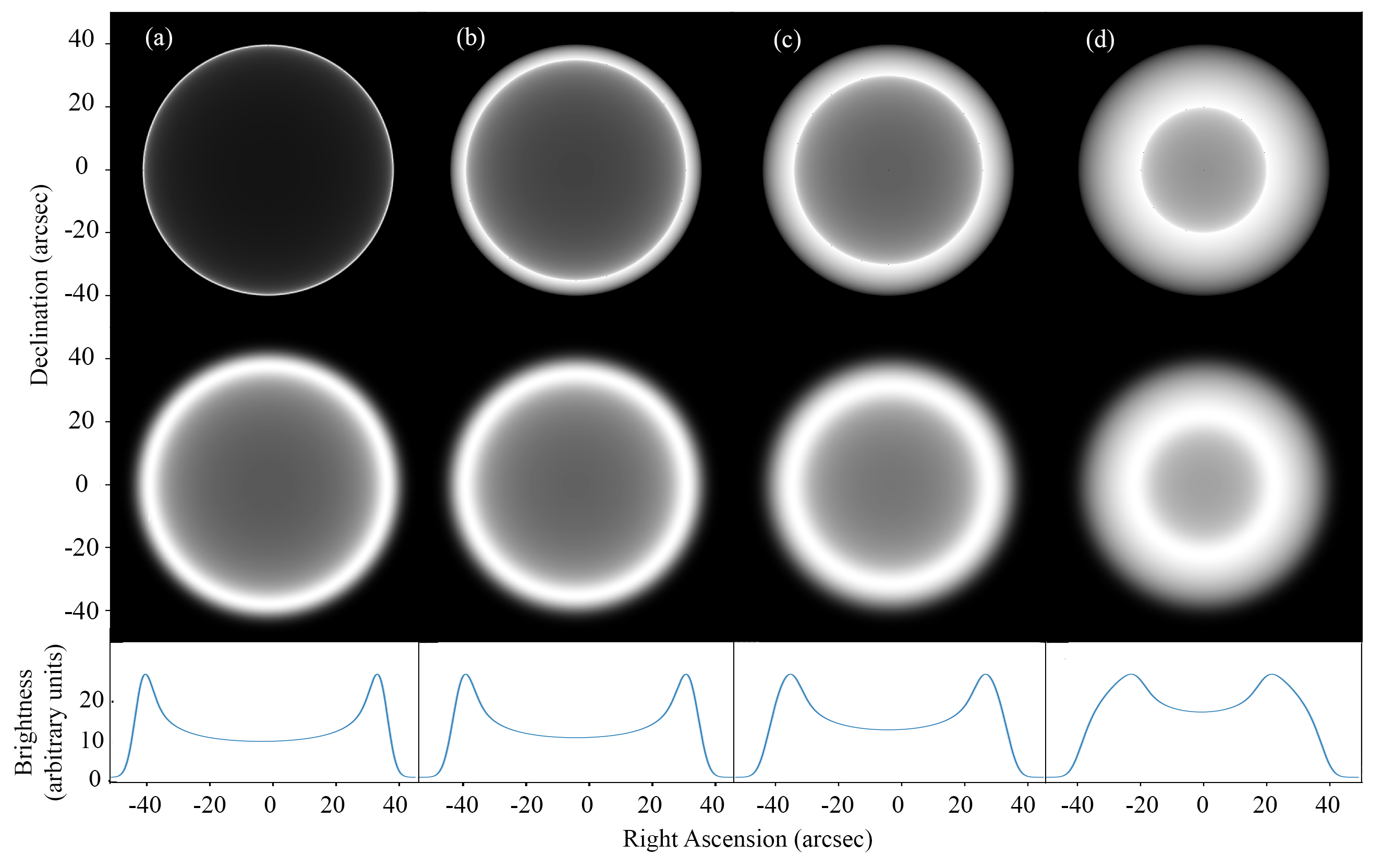}
\caption{Simulation of the appearance of an optically-thin spherical shell of emission for varying shell thickness. In each case, the outer radius of the shell is 40 arcsec, and the inner radii are (a) 39.5, (b) 35, (c) 30, and (d) 20 arcsec respectively. The top row shows the appearance of the model shell, and the second row shows the appearance after convolving with a 6 arcsec FHWM Gaussian beam. The bottom row shows the intensity profile of the convolved image across the diameter cross-section.}
\label{simulation}
%\end{center}
\end{figure*}

For a spherical shell of radius $r$ and thickness $a \ll r$, simple geometry, 
%shown in Figure \ref{geometry}, 
shows that the ratio of the brightness of the limb to the brightness at the center of the shell is $\sqrt{4r/a}$. However, this simple calculation does not take into account the finite resolution of the observation, and so we simulate this effect in Figure \ref{simulation}. The effect of the convolution is that, even with a very thin shell, the ratio of the peak brightness at the limb to the brightness in the centre never exceeds $\sim$ 3.

As predicted by this model, most of the interior of the ORC has a surface brightness of $\gtrsim$ 10 \ujybm, which is within a factor of 3 of the typical surface brightness ($\sim$ 30 \ujybm) at the limb of the ORC.  However,  there is a ``hole'' in the north of the ORC (at  21:03:00.7 -51:59:55) where the surface brightness falls to less than 1 \ujybm, which is a factor $\sim$ 30 below the brightness of the limb. This is inconsistent with the simple spherical shell model, but might be caused by inhomogeneities in the shell.

We measure a minimum thickness of the shell in the image as $\sim$ 7 arcsec, implying a deconvolved thickness of $\sim$ 3-4 arcsec.

% \begin{figure}
% %\begin{center}
% \includegraphics[width=6cm, angle=0]{Figures/geometry.png}
% \caption{If the ORC is a spherical shell of emission of radius $r$, and thickness $a$, then the path length through the limb is $\sqrt(4ar)$, so that the ratio of the surface brightness at the limb to that in the centre is $\sqrt(4r/a)$. However, this ratio is reduced by convolution with the beam of the telescope, as shown in Figure \ref{simulation}.}
% \label{geometry}
% %\end{center}
% \end{figure}

The arc-like internal structures seen in ORC1 are not naturally explained by this model. We now consider the  several potential explanations for these.

\subsection{Equatorial ring}  Figure \ref{interpretations} (a) shows the morphological interpretation in which the outer ring is the limb of a spherical cavity, as discussed above, together with an ``equatorial'' ring.  We note that, on a much smaller scale, the SN1987A supernova remnant shows an elliptical structure very similar to the ``equatorial ring'' shown in Figure \ref{interpretations}(a), and which has been attributed to structure in the circumstellar medium \citep{sn1987a}. Similarly, here we could ascribe this structures to an imhomogeneity in the intergalactic medium. However, without further constraints is it difficult to test or quantify this model.

\subsection{Two ellipses}
\label{ellipses}
Another potential explanation, shown in Figure \ref{interpretations} (b), is that the image consists of two ellipses, caused by a  cylindrical or biconical outflow from the central source, or perhaps from a precessing radio jet that describes a circle as it traverses the interior of a bubble.

Such elliptical structures are well-known in stellar outflows \citep[e.g.][]{akashi16}, and are thought to be responsible for the three rings seen in SN1987A \citep{burrows95}. Biconical outflows have also been seen in galaxies \citep[e.g.][]{veilleux01}, but rings from such outflows  have not previously been seen in an extragalactic source.

\begin{figure}
%\begin{center}
\includegraphics[width=8cm, angle=0]{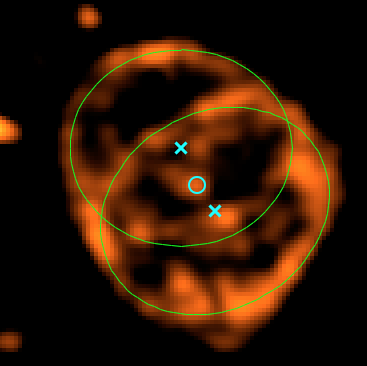}
\caption{Two ellipses fitted to the Meerkat image. The centre of each ellipse is marked with an X.}
\label{fig:ellipses}
%\end{center}
\end{figure}

In Figure \ref{fig:ellipses} we show two ellipses fitted to the Meerkat data by eye, with their centres marked with an X. The two centres form a line with the putative host and are roughly equally spaced, as expected from a cylindrical structure with brightened ends, centred on the host,  and seen at a slight angle to the line of sight. Also one ellipse is slightly larger than the other.

\begin{figure}
%\begin{center}
\includegraphics[width=8cm, angle=0]{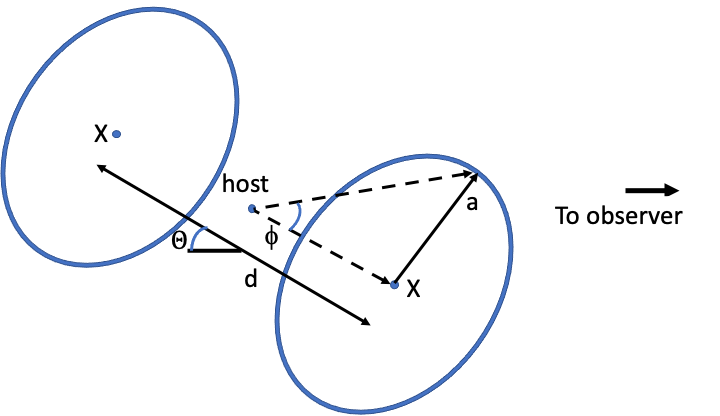}
\caption{A model of the two ellipses shown in Figure \ref{fig:ellipses}, that might be caused by a cylindrical or bipolar outflow from the host, or possibly a precessing radio jet.
}
\label{ellipse-model}
%\end{center}
\end{figure}

We can quantify this model as follows.
Consider a cylinder, shown in Figure \ref{ellipse-model},
%(i.e. could be a biconical outflow) 
of length $d$, radius $a$, tilted at $\theta$ to the line of sight, and that we see the two ends as the two ellipses of the same size. The ellipses will have major and minor axes of $a$ and $a\cos\theta$. The eccentricity is $\sin\theta$, and the observed angular distance between the two centres is $d\sin\theta$.

Fitting this model to the ORC1 data, we obtain $\theta \sim 38\degr$, and the length of the cylinder is $\sim$ 470 kpc. %So the cylinder is quite flattened.
If it is a biconical outflow, then the half-angle of the outflow is $\phi \sim 47\degr$.

The two ellipses have slightly different sizes, and we now consider whether this could be due to expansion during the light-travel time from the back ring to the front ring.
In this case, the nearer ring is $d\cos\theta$ = 370 kpc = 1.2$\times10^6$ light-years closer than the far ring.
From Figure \ref{fig:ellipses}, we estimate the north ring is about 80\% of the size of the south ring. The ring has therefore expanded by 3.4 $\times10^5$ light years in 1.2$\times10^6$ years, so its apparent expansion speed is $\sim$0.28c. 
If this is caused by a spherical shell intersecting a flat screen, then the expansion speed of the shell is  $0.28c\sin\phi = 0.25c $.
We do not know of any mechanism that might produce this expansion speed, so we conclude that the difference in the size of the two rings is due to an intrinsic effect rather than light-travel time.

\subsection{Other structure}
None of the above interpretations naturally explains the other faint structure seen in the ring, as shown in Figure \ref{interpretations}(c). Are these faint remnants of previous activity, or do they indicate yet another mechanism  such as the galaxy trails discussed  in Section \ref{environment}?

\subsection{Comparison with Other ORCs}
\label{comparison}
\begin{figure*}
%\begin{center}
\includegraphics[width=18cm, angle=0]{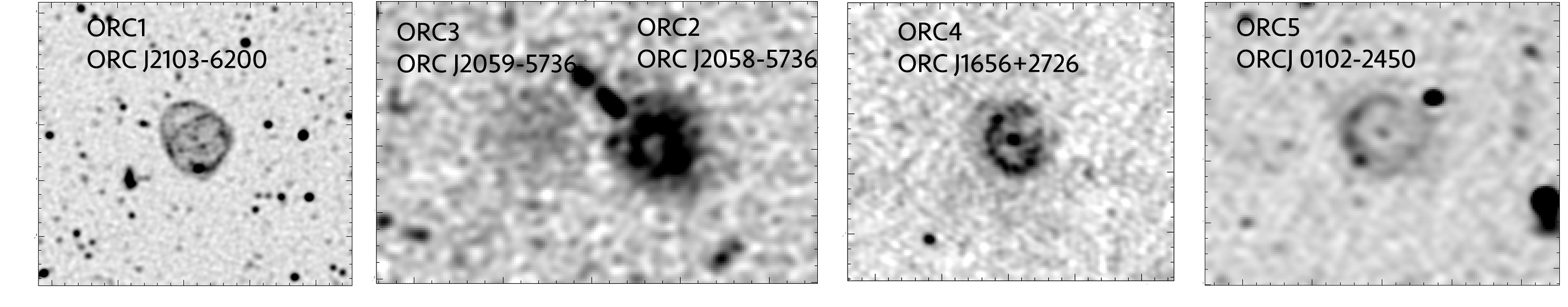}
\caption{Radio images of the five published ORCs, each at a similar scale. The image of ORC1 is from MeerKAT observations described in this paper, ORC2 and 3 are from ASKAP observations \citep{norris20}, ORC4 is from GMRT data \citep{norris20}, and ORC5 is from ASKAP data \citep{koribalski21}.
}
\label{allorcs}
%\end{center}
\end{figure*}

Radio images of the five published ORCs are shown in Figure \ref{allorcs}, and a summary of their properties is given in Table \ref{tab:allorcs}. We consider the photometric redshifts  reliable because (a) \citet{norris21} compared the \citet{zou19, zou20} results with spectroscopic redshifts where available, and found good agreement, and (b) we have independently estimated the photometric redshifts for the galaxies in Table \ref{allorcs}, using a kNN algorithm \citep{luken19} completely different from that used by Zou et al., and find good agreement in the redshift values.

\begin{table*}
\centering
%\tiny
\caption{A summary of the properties of the five published ORCs. The redshift in each case is the photometric redshift of the central galaxy \citep{zou19, zou20}. Radio flux density includes sources within the ORC.}
\setlength{\tabcolsep}{3pt} % reduces space between columns to fit on the page horizontally
\renewcommand{\arraystretch}{1.25} % increases space between rows to make it more readable
\label{tab:allorcs}
\begin{tabular}{ccccccccccccccccl}
\hline
Short & Full & assumed & diameter & diameter & Integrated\\
name  & name & redshift & (arcsec) & (kpc) & radio flux density\\
& & & & & (mJy at 1 GHz) \\
\hline
ORC1 & ORC J2103-6200  & 0.551 & 80 & 520 & 3.9\\
ORC2 & ORC J2058-5736b & ? & 80 & & 4.9\\
ORC3 & ORC J2059-5736a & ? & 80 & & 1.7\\
ORC4 & ORC J1656+2726  & 0.385 & 90 & 400 & 9.4\\
ORC5 & ORC J0102-2450  & 0.27 & 70 & 300 & 3.2\\
\hline
\end{tabular}
\end{table*}

Of the five published ORCs, two (ORC J2058-5736b = ORC2 and ORC J2059-5736a = ORC3) are a pair, quite unlike the remaining three ORCs, which are all single.  They are also different in their radio morphology. The three remaining ORCs each have an edge-brightened ring structure, with diffuse emission in the centre, and little diffuse emission outside the ring. By contrast, ORC2 has a disc of diffuse emission which extends significantly further than the ring, and ORC3 is a roughly uniform disc, rather than being edge-brightened.

Because of these differences, we regard the pair of ORC2 and ORC3 as being significantly different from the other three ORCs, and in the rest of this paper we will treat the three single ORCs as one class, and the double ORC (ORCs 2 and 3) as a separate class. 

Each of the three single ORCs has a central galaxy which is detected at radio, infrared, and optical wavelengths. The multiwavelength properties of the  central galaxies are summarised in Table \ref{tab:optical}.

The density of radio sources $> 40$ \ujy\ at 1284 MHz is about 1980 per square degree. Each of the three single ORCs has a  galaxy with radio emission at least as strong as this  within 10 arcsec of the centre, and so the probability of one of these galaxies being a chance coincidence is $\sim$ 0.05. The probability of all three having such a galaxy by chance is therefore $\sim 10^{-4}$. We therefore consider it unlikely that the central galaxies are a  chance alignment, and instead consider that they must be an important part of the mechanism driving the ORCs.

Hereafter we refer to these central galaxies as the ``host'' galaxies of the ORCs.

\subsection{Space Density of ORCs}
\label{density}
The observed space density of ORCs is very uncertain, both because of the small sample size, and because of the uncertainty in the definition of what constitutes an ORC (e.g. we are investigating several other unpublished candidate ORCs). Nevertheless, we  estimate the observed density of ORCs in ASKAP data, for the purposes of this discussion, to be about 1 ORC per 50 square degrees.
 
All three single ORCs have a similar size ($\sim$ 1 arcmin diameter) and redshift ($z \sim 0.2 - 0.6$). This similarity could be attributed to (a) specific physical conditions that favour this size and redshift range, (b) pure coincidence due to the small sample size, or (c) a selection effect. 
 
For example, if all ORCs were physically $\sim$ 1 Mpc in diameter, it is possible that nearby ORCs would be too extended to detect, while more distant ORCs might fall below the EMU surface brightness limit, since surface brightness decreases in a non-Euclidean Universe as $(1+z)^4$ \citep{tolman34,calvi14}. This seems unlikely to be a strong constraint, because (a) EMU is sensitive to a wide range of angular scale sizes \citep[see Figure 18 of][]{norris21}, and (b) the ORCs are well above the sensitivity limit.
 
If we assume that we can only detect ORCs in the redshift range 0.2 < z < 0.6, then the observed comoving volume density is 1 ORC per 0.05 Gpc$^{3}$, or $2\times10^{-8}$ per Mpc$^3$. This is comparable with the space density (10$^{-6.8}$ Mpc$^{-3}$) of starburst galaxies with SFR > 100 \Msun/yr as inferred from
 \citet{bothwell11}.

\section{Potential Causes of the ORC phenomenon}
In this section, we consider the mechanisms that might cause the ORC phenomenon. We take as our starting point that we observe a circular edge-brightened region of diffuse emission, with possible interpretations as discussed in Section \ref{morphology} above,   surrounding a galaxy with properties as described in Section \ref{galaxy} above. 

This ring is most readily explained in terms of a shell of synchrotron emission, and so we first consider the implications of that explanation.  We then  consider three possible physical models (an explosion in the host galaxy, an end-on radio lobe, and a starburst termination shock). Here we consider the starburst termination shock in some detail, and will consider the other models in detail in future papers. Derivations of the equations used below for synchrotron radiation may be found in \citet{dummies}.

\subsection{General considerations}

A completely successful model needs to explain the following observational properties: 
\begin{enumerate}

\item the approximately circular, edge-brightened structure; 

\item the arc sub-structures;

\item 
The well-ordered magnetic field revealed by the polarisation phenomenology; 

\item %(given a redshift or distance determination)
the total radio luminosity and the total non-thermal (i.e., cosmic ray + magnetic field) energy content;

\item %(given a redshift or distance determination)
the physical dimensions of the structure, including the ring radius $r$, ring thickness $a$, and  the radial intensity profile; 

\item the non-thermal radio spectrum, with an observed spectral index $\sim$ -1.6  
over the frequency range of $\sim$0.1 -- 2 GHz;

\item the observed comoving volume density of ORCs, which is roughly 1 ORC per 0.05 Gpc$^{3}$, or $2\times10^{-8}$ per Mpc$^3$.

\end{enumerate}

\subsection {Synchrotron emission in a Spherical Shell}
\label{synchrotron}
Several models (e.g. blast wave, starburst outflow) explain the observed rings as being  a spherical shell  of synchrotron emission, and here we explore the properties of this putative shell, without regard to the underlying mechanism that drives it. 
We assume the synchrotron electrons have been accelerated by a spherical shock from the host galaxy. 

We do not have a reliable guide to the magnetic field around the shock. While ambient magnetic fields as high as 40 $\mu$G have been reported in clusters \citep{carilli02}, \citet{govoni19} find a magnetic field in the ridge at z$\sim$0.07 connecting two clusters to be  < 1$\mu$G. We showed in Section \ref{environment} that ORC1C lies in a significant over-density, and so an ambient magnetic field of a few  $\mu$G seems reasonable. 
If the ORC shell is produced by a shock, then this ambient field may then be amplified by the shock by a factor of a few. 
This is comparable to the B field in equipartition with the CMB (cosmic microwave background) at z = 0.55 of 8 $\mu$G, which is  further discussed below.

In summary, we do not have a good measure of the magnetic field, but in the next Section we derive an estimate based on the observed shell thickness.

\subsubsection{Spectral Index and Electron energy distribution}
The steep spectral index suggests synchrotron radiation by a non-thermal electron distribution in an ambient magnetic field. 

For cooled electrons,
this implies a power-law electron energy distribution 
$dN_e/dE_e \propto E_e^{\gamma_{\rm e}}$ 
with spectral index $\gamma_{\rm e}$ related to the observed spectral index $\alpha$ as $\gamma_{\rm e} = (2 \alpha - 1) = -3.82$
covering the minimum range in electron energy $E_e$ of $\sim (3.1 - 14)$ GeV $(B/\mu G)^{-1/2}$, where $B$ is the magnetic field.

This  steep distribution (i.e. low value of $\gamma_{\rm e}$) suggests that the electron distribution is in steady-state and cooled by either the synchrotron process itself or inverse Compton (IC) cooling by  the 
CMB.

\begin{figure}
%\begin{center}
\includegraphics[width=8cm, angle=0]{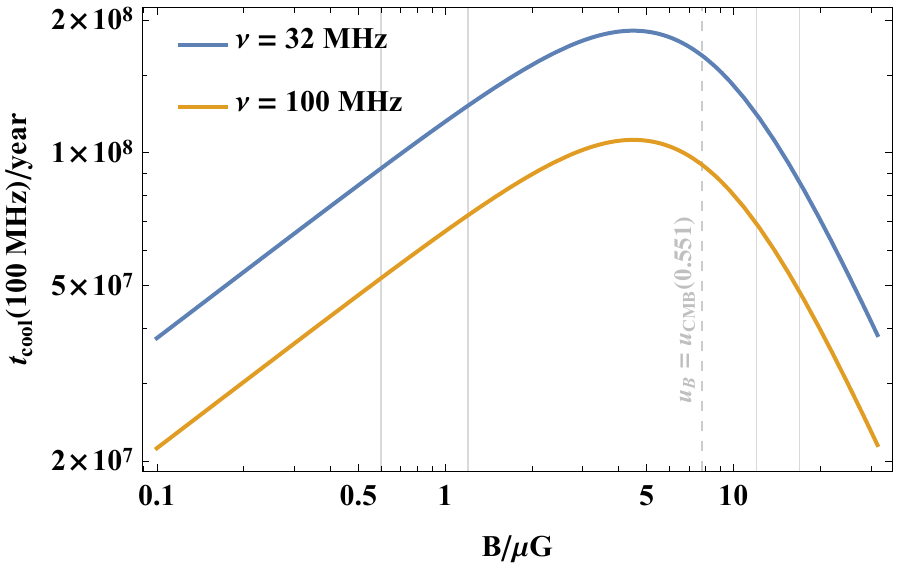}
\caption{The cooling time of electrons synchrotron-radiating at 100 MHz (the minimum frequency for which we have a flux density measurement) and (for illustrative purposes) at 32 MHz. 
For $B \gtrsim 8\mu G$, the cooling is dominated by synchrotron radiation, but below this value it is dominated by Inverse Compton (IC) cooling, whose magnitude  is determined by the energy density in the CMB, $\propto (1+z)^4$. 
The vertical dashed line shows the magnetic field amplitude in energy partition with the CMB at $z = 0.551$.
The other vertical lines show the two, separated ranges of magnetic field amplitude that reproduce the inferred thickness of the radio-emitting rim of ORC1 (cf.~\autoref{fig:emitThickness}).
} 
\label{fig:tcool}
%\end{center}
\end{figure}

Figure.~\ref{fig:tcool} shows that the cooling timescale is $\sim$ 10$^8$ yr),
%assuming a $\sim \mu$G field. This is comparable
which is comparable to the transport timescale between ORC1C and the ORC
rim at a fiducial speed of 1000 km/s, 
approximately corresponding to the sound speed in a galaxy cluster or to the expected
speed of a star-formation-driven wind:
\begin{equation}
    t_{\rm adv} \equiv r/v \simeq 260 \ {\rm kpc}/v \simeq 2.6 \times 10^8 \ {\rm year}/v_3
\label{eq:tadv}
\end{equation}
where $v_3 \equiv v/(10^3 \ {\rm km/s})$.

Thus, either the electrons have not been re-accelerated for at least the past $\sim$ 100 Myr, 
or, in the case of  continuous acceleration,  the electrons are transported out of the acceleration zone (presumably the vicinity of a shock) and subsequently cool via synchrotron and possibly IC for $\gtrsim$ 100 Myr.

This scenario of synchrotron emission from a cooled electron distribution implies that the
central value of the {\it injection} spectral index of the electron distribution is
$\gamma_{\rm e, inj} = \gamma_{\rm e} + 1 = -3.82 + 1 = -2.82$.
This suggests diffusive acceleration at a shock or shocks of  Mach number $\mathcal{M} \sim 2-3$ using the standard theory \citep{Bell1978} of diffusive shock acceleration,
in the test particle limit.

However, we note several caveats to such conclusions founded on the measured spectral index.
\begin{itemize}
    \item Recent observations \citep{Rajpurohit2020} 
show that the radio spectra of the `Toothbrush Relic' shock front
 follow near-perfect power laws, with
spectral indices close to the relic's 
integrated spectral index,
$-1.16 \pm 0.02$ (which is close to the ORC1 integrated spectral index).
However, simulations \citep{Wittor2019} demonstrate that this apparent uniformity can  result from averaging over reacceleration processes occurring in multiple shock sub-zones.
\item The standard relation between spectral index and shock Mach number obtained in the test particle limit will not necessarily hold in the case that the 
observed synchrotron emission is not due 
primarily to a  CR electron population
accelerated up out of the thermal pool at the 
shock, but rather due to a pre-existing `fossil' non-thermal electron population {\it re}accelerated at the shock.
Such a situation has been claimed by \citet{Colafrancesco2017} for  cluster radio relics where X-ray and radio spectral index observations point to inconsistent Mach numbers.
A pre-existing non-thermal population available for reacceleration may also render the shock apparently surprisingly efficient in converting mechanical power into non-thermal particles, even in the case it is characterised by a low Mach number.

\end{itemize}

\begin{figure}
%\begin{center}
\includegraphics[width=8cm, angle=0]{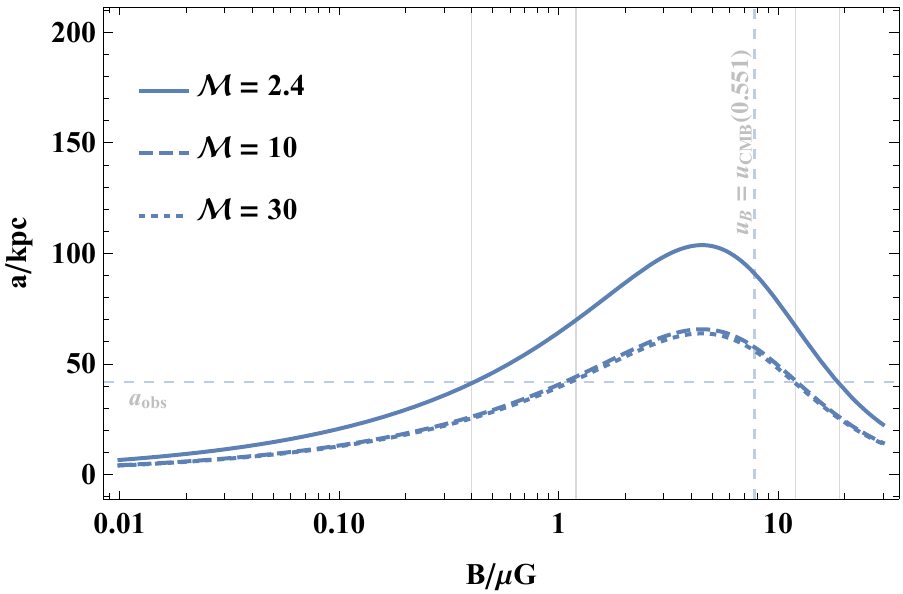}
\caption{The physical length scale over which 1 GHz synchrotron emission from electrons, accelerated at the putative shock at the inner edge of the ORC rim and radiating in the specified magnetic field, drops to 30\% of its initial intensity  \citep{Hoeft2007}.
The vertical dashed line marks where a magnetic field of the specified amplitude is in equipartition with the CMB energy density for the redshift of ORC1C. 
The horizontal line shows the linear scale $\simeq 40$ kpc
corresponding to $6.5''$, the  angular scale over which the intensity is measured to drop to 30\% of its peak value; the solid vertical lines delineate the two separate regions of
magnetic field parameter space where, for some value of the somewhat uncertain shock Mach number, the predicted thickness of the synchrotron emission region matches that of the observed radio shell.
}
\label{fig:emitThickness}
%\end{center}
\end{figure}

This model implies that this electron population will cool by synchrotron and IC emission as it is transported
from the region of the shock.
The combination of downstream flow velocity and cooling time determines the width of the downstream synchrotron-emission zone and hence the shell thickness \cite[e.g.,][]{Hoeft2007}.
Thus, 
we can calculate the range of magnetic field amplitude for which the model and observed shell thicknesses agree. Specifically, we use the z-dependent distance inferred from the angular size over which the intensity drops to 30\% of its peak value, viz.~$6.5''$).
This calculation is shown in \autoref{fig:emitThickness}.

Qualitatively, the behaviour of these curves can be understood as follows: for relatively small magnetic fields well to the left of the vertical dashed line (i.e., for $u_B \ll u_{\rm CMB}(z)$, where $u_B$ and $u_{\rm CMB}(z)$ are the magnetic field and redshift-dependent CMB energy density, respectively), IC losses set the cooling time of the electrons and the range of the electrons that synchrotron-radiate at a specified, fixed frequency (1 GHz) increases with $B$ (because increasing $B$ lowers the energy implied for the emitting electrons, thus increasing their IC loss time). 
Once equipartition is approached and then exceeded (to the right of the dashed vertical line), synchrotron losses themselves become the dominant process setting the cooling time, and thereafter
the range starts to decline with $B$.

Unfortunately, this behaviour means that there are two $B$ ranges where the observed and modelled shell thickness match.
To break this degeneracy, we 
consider the energetics implied by the measured radio flux density.
The CR electron population and the large scale magnetic field each  represent an energy reservoir, and we can ask:
For what (mean) magnetic field is the total energy in CRe's + magnetic field minimised?

\begin{figure}
%\begin{center}
\includegraphics[width=8cm, angle=0]{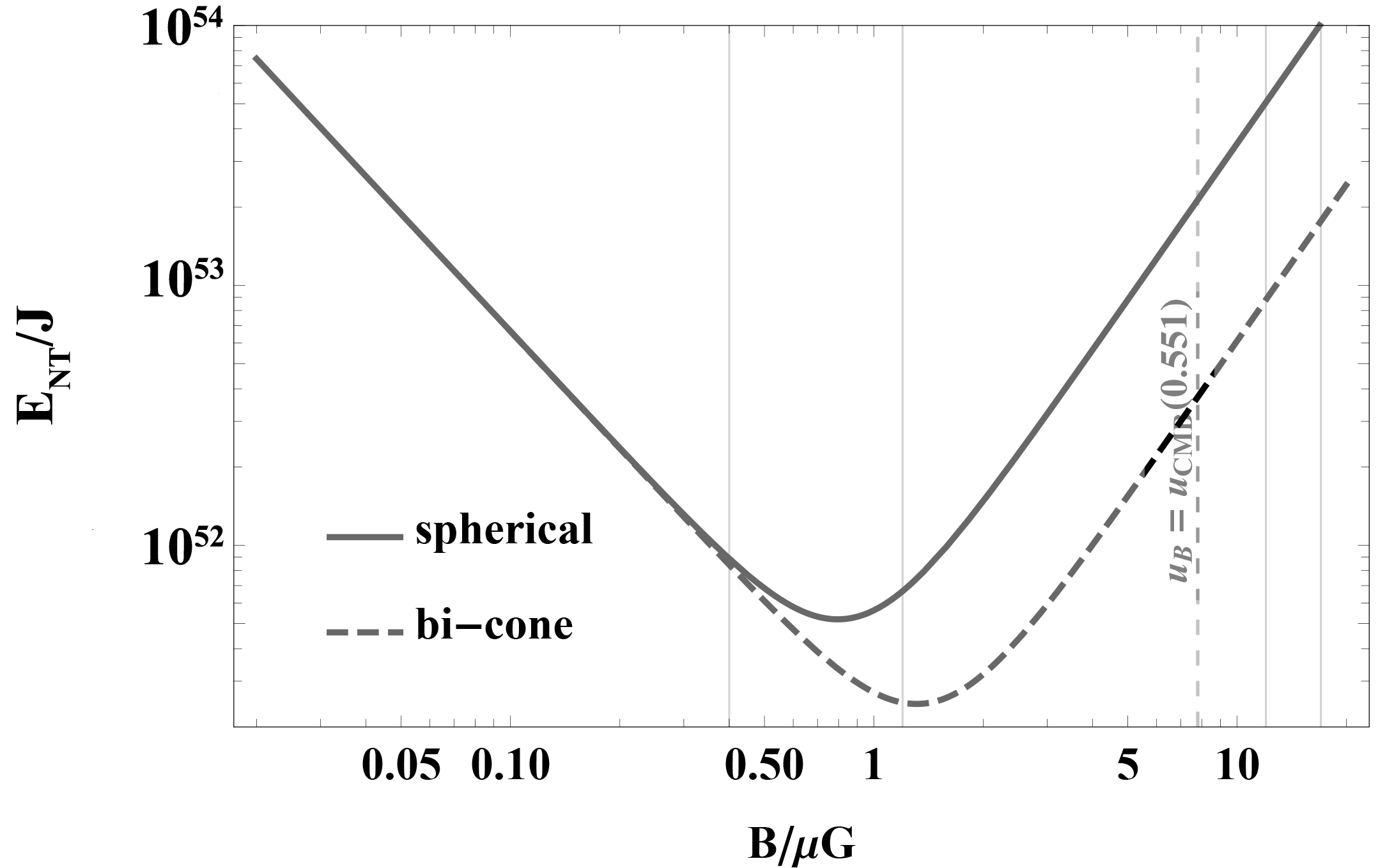}
\caption{The total non-thermal `NT' = CR electron + magnetic field energy of ORC1 as a function of
magnetic field amplitude $B$ as measured from its  radio brightness.
The two solid vertical lines span the lower of the two regions of magnetic field intensity favoured by the analysis presented in \autoref{fig:emitThickness}.
For the baseline `spherical' case we assume that the shell of enhanced magnetic field is 40 pc in extent, with an outer radius of 260 kpc (the alternative `bi-cone' case is for the geometrical interpretation shown in \autoref{interpretations}, panel c.)
} 
\label{fig:plotMinEnergyConfig}
%\end{center}
\end{figure}

The radio luminosity of  ORC1 is given by
\begin{eqnarray}
    L_{\rm radio} & = & 4 \pi d^2_L(z) 
    (1 + z)^{-(1 + \alpha)} 
    \int_{\nu_{\rm min}}^{\nu_{\rm max}} d \nu \ F_\nu(\nu) \nonumber \\
%    & \simeq & 4.9 \times 10^{41} \  {\rm erg/s}
    & \simeq & 4.9 \times 10^{34} \  {\rm W}  \ ,
\label{eq:Lradio}
\end{eqnarray}
However, the observed radio luminosity 
is equal to
the total radiated power only in the limit $u_B \gg u_{\rm CMB}$; in general we should correct \autoref{eq:Lradio}
to include the inferred IC luminosity:
\begin{eqnarray}
    L_{\rm tot} & \equiv & L_{\rm radio} + L_{\rm IC} \nonumber \\
    & \simeq & \left(\frac{u_B + u_{\rm CMB}}{u_B}\right) L_{\rm radio} \nonumber \\
%    & \simeq & 4.9 \times 10^{41} \  \left(1 + \left(\frac{B}{\rm 8 \ \mu G}\right)^{-2} \right) {\rm erg/s}  \ ,
    & \simeq & 4.9 \times 10^{34} \  \left(1 + B_{\rm CMB}^{-2} \right) {\rm W}  \ ,
%    & \simeq & 4.9 \times 10^{41} \  \left(1 + B_{\rm CMB}^{-2} \right) {\rm erg/s}  \ ,
\label{eq:Ltot}
\end{eqnarray}
where $B_{\rm CMB} \simeq B/(\rm 7.8 \ \mu G)$, is the magnetic field amplitude in units of the magnetic field that is in energy density equipartition with the CMB at $z=0.551$;
$L_{\rm tot}$ calculated according to \autoref{eq:Ltot} is displayed in
\autoref{fig:plotMinEnergyConfig}.

Given $ L_{\rm radio}$ and a magnetic field amplitude $B$, and
assuming a quasi steady state, the total energy in the CR electrons is:
\begin{equation}
    E_e(B) \simeq \frac{L_{\rm radio}}{t_{\rm synch}(B)},
\end{equation}
where $t_{\rm synch}(B)$ is the synchrotron loss time for electrons radiating 
in a magnetic field $B$.
The total non-thermal (CR electron + magnetic field) energy of the ORC as a function of magnetic field amplitude
$E_{\rm NT}(B)  \equiv  E_e(B) + E_B(B) = E_e(B) + u_B(B) v_{\rm shell}$
where $v_{\rm shell} \simeq 3.0 \times 10^7$ kpc$^3$ is the volume of the 40 pc thick emitting shell and $u_B$ is the magnetic field energy density.

The result of this calculation is shown in \autoref{fig:plotMinEnergyConfig}.
It is evident from this figure that the upper magnetic field region favoured by the shell thickness analysis presented in \autoref{fig:emitThickness} would require a total non-thermal energy content around two orders of magnitude larger than that required in the lower region.
On this basis we firmly favour a magnetic field $\sim$ 0.8 $\mu$G.
While it is not necessary, of course, that the system actually be in a configuration of minimum total non-thermal energy, it is suggestive that such a  configuration would possess a magnetic field amplitude squarely in the middle of the lower range given by the shell thickness analysis.
Adopting a 0.8 $\mu$G field, we determine that the energy density in CR electrons (+ positrons if present) in the synchrotron emitting shell is $\sim 2.1 \times 10^{4}$ eV m$^{-3}$, around 1\% of the energy density in the CMB at $z = 0.551$.

%}

\begin{figure}
%\begin{center}
\includegraphics[width=8cm, angle=0]{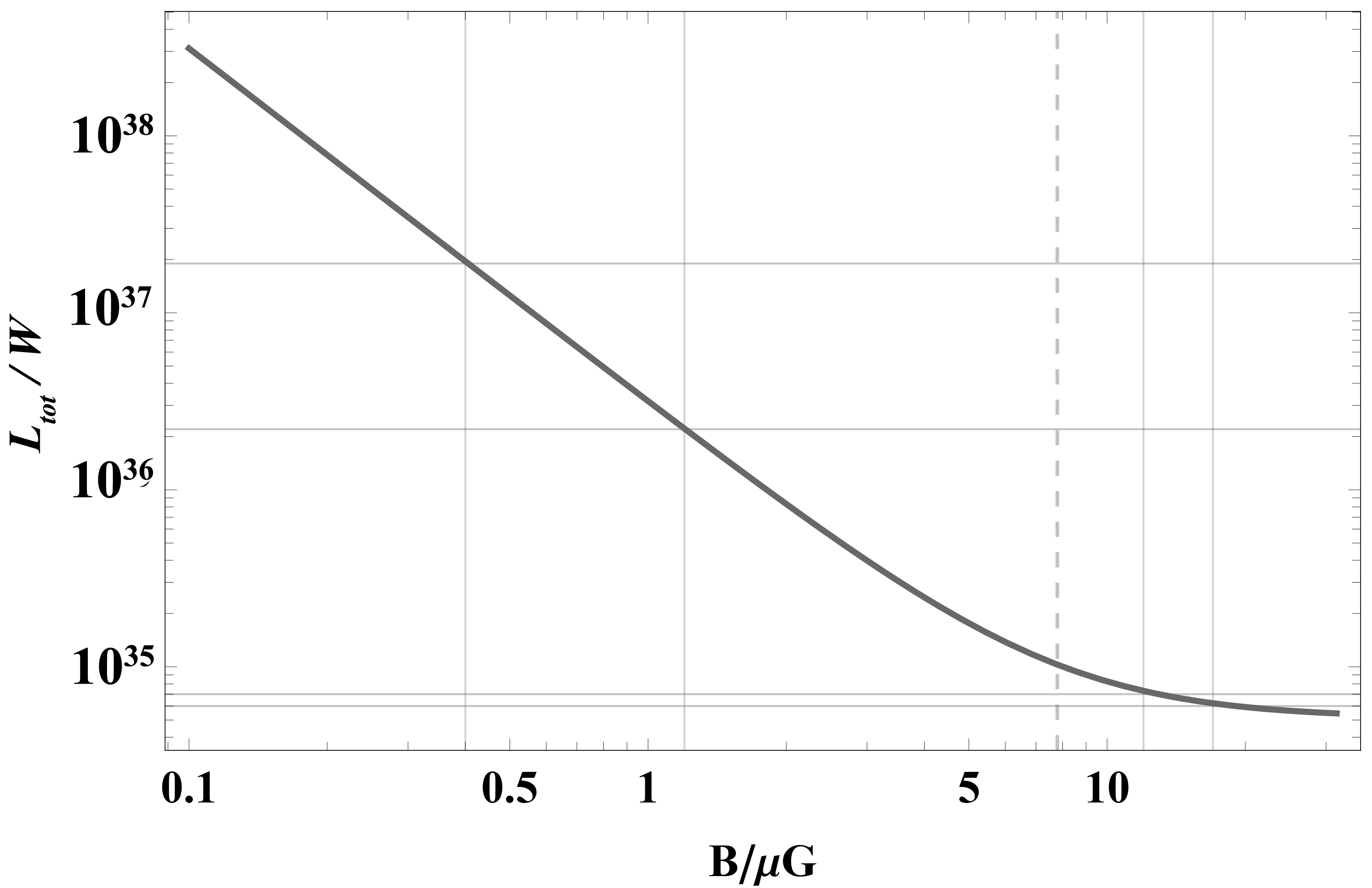}
\caption{The total non-thermal luminosity (synchrotron + IC) of CR electrons (+ positrons) in the ORC as a function
of ambient magnetic field.
The vertical solid and dashed lines have the same meanings as established in \autoref{fig:emitThickness}; 
in the low field region, the total non-thermal luminosity is in the range $(2-12) \times 10^{36}$ W; in the high field region it is in the range
$(6-7) \times 10^{34}$ W.
} 
\label{fig:plotOCRLuminosity}
%\end{center}
\end{figure}

Note that the IC signal predicted in the case  $B \sim 0.8 \ \mu$G regime emerges (for the $\sim 10$ GeV CR electrons synchrotron radiating at 1 GHz) 
at a rest-frame energy of $\sim$440 keV, but this is redshifted down to $\sim$280 keV, placing it into the very hard X-ray band; the photon number flux of this signal at that energy should be around
$3 \times 10^{-4}$ m$^{-2}$ s$^{-1}$.

A further prediction of this model is that we expect to see an age gradient of the electrons across the shell, which should result in a spectral index gradient across the shell, with the outside of the shell, containing newly-shocked electrons,  having a flatter spectral index than the ageing electrons on the  inside of the shell. Unfortunately our spectral index data do not have enough spatial resolution to show this, but we do note that  spectral index in the interior of the ring is steeper than on the ring.

\subsection{Explosion in the host galaxy}
 
Perhaps the most obvious explanation of a spherical shell of synchrotron emission is a spherical shock from a cataclysmic event in the host galaxy \citep{norris21, koribalski21}, such as the merger of two supermassive black holes \citep[e.g.][]{bode12}. The resulting shock wave would accelerate electrons in the inter-galactic medium, resulting in a spherical bubble of radio emission, as described above in Section \ref{synchrotron}, which we would observe as an edge-brightened ring. 

The properties of this shock would be similar to that discussed  in Subsection \ref{starburst}. 
\citet{koribalski21}  estimate an available energy of $\sim 6 \times 10^{53}/M^2$ J (assuming the shock moves at Mach $M \sim 1$), and an age of $\sim 57/M$~Myr. Assuming a kinetic-to-synchrotron power conversion efficiency of $\sim 4 \times 10^{-3}$ as inferred from the FIR-radio correlation for star-forming galaxies, this corresponds to an integrated radio power of $3 \times 10^{38}$~W, and a 1.2 GHz luminosity of $L_{1.2} \approx 4 \times 10^{28} / M$~W\,Hz$^{-1}$ This value is an upper limit: diffusive shock acceleration will become less efficient as the bubble expands, resulting in the observed steepening of the spectrum from the initial $\alpha_{\rm inj} \approx -0.5$ power-law. Models of energy injection in remnant radio galaxy lobes (below) yield $\nu L_\nu / Q \approx 10^{-5} - 10^{-6}$ for lobes at equipartition; if this efficiency is adopted for the shell model, integrated 1 GHz luminosities of order $\sim 10^{25}$ ~W\,Hz$^{-1}$ would be reasonably expected, consistent with the observed integrated flux and a source at $z=0.551$.   

If ORCs are the result of a merger of SMBH, we would expect ORC hosts  to contain a  SMBH, which may be manifested as an AGN. We note that:
\begin{itemize}
    \item the radio emission from the ORC1 host galaxy is higher than can be accounted for by the current low SFR, and so must be due to an AGN,
    \item the radio emission from the ORC4 host galaxy corresponds to a SFR of $\sim$200 \Msun/yr which, while physically possible, is very rare, making it likely that this is due to an AGN, and
    \item \cite{koribalski21} argue that the host of ORC5 must contain a SMBH which is responsible for the radio emission from that galaxy. 
\end{itemize}

Thus all three hosts of the single ORCs contains a radio-loud AGN, which is consistent with this hypothesis that ORCS are caused by a merger of SMBH.

\subsection{End-on Radio lobes}
Radio-loud AGN often consist of two jets of relativistic plasma ejected from the host galaxy and cooling into two diffuse lobes of radio emission.
Here we consider whether ORCs may result from viewing such a system end-on, so that the two lobes appear superimposed, with the host galaxy appearing at the centre.

As noted in the previous subsection, all three hosts of the single ORCs contains a radio-loud AGN, which in principle could generate the required radio lobes.

However, this model needs to overcome three hurdles:
\begin{itemize}
    \item In such a system, the jet would be directed close to our line of sight, so would appear as a quasar or blazar, with the central source much brighter than the diffuse lobes. This hurdle can be overcome if the central AGN has switched off, leaving the lobes which have a much longer decay time.  % relic lobes
    \item If the jet is oriented at an angle $\theta$ to the line of sight, the number of such systems should be proportional to $\sin^2\theta$ so that many more sources should be seen with the front and back lobes slightly displaced than are seen with the front and back lobes aligned, resulting in many more ``double ORCs'' or ``figure of 8'' galaxies than circular ORCs, but observations so far have found 3 single ORCs \citep{norris21, koribalski21} but only one double ORC \citep{norris21}. 
    \item The diffuse lobes of AGN tend to be very irregular, so that an additional mechanism is needed to produce the remarkable circular ring, with a well-defined narrow rim,  seen in the ORCs.
Mechanisms such as vortex rings were suggested by \citet{koribalski21} to overcome this hurdle. An alternative solution is that the plasma from a jet that has switched off may rise buoyantly in the extragalactic medium, producing circular rings similar to smoke rings. The challenge for this latter interpretation is that remnant lobes fade rapidly, while the dynamical transformation of lobes into toroidal bubbles occurs on longer timescales.
\end{itemize}

Detailed 3-D simulations aimed at addressing these questions are currently underway (Shabala et al., in preparation).

\subsection{Starburst termination shock}
\label{starburst}
Even a
modest star-forming 
galaxy, like the Milky Way, is expected to launch a galactic wind.
For a galaxy located outside 
a cluster environment, this wind 
can be expected to form a termination shock at a distance of a few $\times$ 100 kpc \citep{jokipii87, volk04, bustard17, lochhaas18}. 
At such large distances from the host, the wind into each galactic hemisphere may expand radially, resulting in an approximately spherical shock structure, assuming an isotropic  environment. 
The wind termination shock can accelerate electrons into non-thermal distributions
at $\sim$ 10 GeV energies \citep[][]{bustard17} 
which can then produce synchrotron radiation at $\sim$ 1 GHz.

The power in such winds is roughly proportional to the star formation rate \citep[e.g.][]{strickland00,lochhaas18}, and is dominated by a
hot ($T \gtrsim 10^6$ K at launch) phase with an asymptotic speed of v$_{\rm wind} \gtrsim 1000$ km/s.
The  total energy available to launch a wind by 
the formation of a mass $M_\star$
of 
stars  is \cite[e.g.,][]{Crocker2021}:
\begin{eqnarray}
    E_{\rm mech, SF} & \simeq &
   {\rm 10^{44} \ J} \ \left(\frac{M_\star}{\rm 95 \ M_\odot} \right),
\label{eq:EMech} 
\end{eqnarray}
and the instantaneous mechanical power injected by galactic star formation can therefore be estimated as
\begin{eqnarray}
    \dot{E}_{\rm mech, SF} & \simeq &{\rm SFR} \times  \left(\frac{\rm 10^{44} \ J}{\rm 95 \ M_\odot}\right) \nonumber \\
 & \simeq & 6.3 \times 10^{35} \  {\rm W}  \ \left(\frac{\rm SFR}{\rm 20 \ M_\odot/yr}\right) ,
\label{eq:EdotMech} 
\end{eqnarray}
We can compare \autoref{eq:EMech} and \autoref{eq:EdotMech}
with  \autoref{fig:plotMinEnergyConfig} and 
\autoref{fig:plotOCRLuminosity}, respectively.
To supply the total non-thermal (magnetic field + CR electron + positron) energy content of the ORC we require the formation of a minimum total mass in stars
\begin{equation}
    M_{\rm \star,ORC} \simeq 2.4 \times 10^{10} \ {\rm M_\odot} \ \left(\frac{\eta_{\rm NT}}{0.2}\right) \left(\frac{E_{\rm NT, ORC}}{5 \times 10^{51} \ {\rm J}}\right)
\label{eq:minStarMass}
\end{equation}
where $\eta_{\rm NT}$ is the efficiency for the conversion
of the mechanical energy released by star formation to the energy contained in magnetic fields and cosmic ray electrons (+ positrons).
Normalising
%, for the moment, 
the total non-thermal (synchrotron + IC) luminosity of the ORC for the low $\sim 1 \mu$G field solution,
the ORC requires
an instantaneous power equivalent to that
released by 
%quasi steady state 
star formation at a rate of 
\begin{equation}
    \dot{M}_{\rm \star,ORC} \simeq 300 \ {\rm M_\odot/yr}
    \ \left(\frac{\epsilon_{\rm e}}{0.2}\right)
    \ \left(\frac{L_{\rm tot}}{\rm 2 \times 10^{36} \ W} \right)
\label{eq:minSFR}
\end{equation}
where $\epsilon_{\rm e}$ is a nominal efficiency for converting
mechanical power released by star formation into 
non-thermal electrons.

Clearly \autoref{eq:minStarMass} and \autoref{eq:minSFR} are extremely demanding constraints.
An immediate problem posed by \autoref{eq:minSFR} is that this SFR is much larger than the current SFR estimates  from either {\sc ProSpect} or {\sc MAGPHYS}
 presented in \autoref{sec:SFR}.
In the case that we can simply assume the large (12-17 $\mu$G) field solution, the power requirements are considerably relieved and might conceivably be satisfied by a SFR of 10-20
\Msun/yr, but this SFR, while potentially in agreement with the SFR estimate for the host galaxy founded purely on its
radio continuum luminosity, is still in considerable tension with the {\sc ProSpect} estimate.
Moreover, the severe problem with a putative high field scenario is the huge magnetic field energy implied (\autoref{eq:EMech}), far larger
than could conceivably by supplied by star formation within ORC1C.
As mentioned above, however, a field in the few $\mu$G range might be produced via extrinsic processes in the case of a sufficiently dense cluster environment but such a strong field conceivably becomes dynamically important, a scenario that requires exploration beyond the scope of this paper.

Instead, we consider the low magnetic field solution, which requires a much higher SFR than supported by Section \ref{sec:SFR}.
However, 
the long advection timescale between the host galaxy and the ORC shell, $\sim 3 \times 10^8$ yr from \autoref{eq:tadv}, implies that the current star-formation activity of ORC1C is less important than its star formation history over the last $\sim$Gyr

Semi-analytic modelling 
by \citet{lochhaas18}  explores the phenomenology of the bubble structures caused by galactic winds from star-forming galaxies in three different SFR regimes (1,10, and 100 \Msun/yr)
in a scenario where a galaxy turns $4 \times 10^9$ \Msun \ of gas %-- equivalent to the  entire gas content of the Milky Way -- 
into stars.

The resulting structures have a  nested structure of (going from inner to outer radii) wind termination (or reverse) shock, contact discontinuity, and forward shock.
\citet{lochhaas18} show a clear time delay effect in the 10 and 100 \Msun/yr cases: the  contact discontinuities continue to expand
out to  radii of a few 100 kpc on timescales of a few Gyr, while the star formation itself ceases well before 1 Gyr.

The size and energy content of the bubbles modelled by \citet{lochhaas18} suggest a
%make them a tantalizing -- if not yet compelling -- 
possible analogue for the ORC phenomenon, to be discussed further in Crocker et al. (in preparation ).

Important  questions remain open here.
In particular, it is still unclear whether the star formation history of ORC1C 
can match the energy and power requirements implied by 
\autoref{eq:minStarMass} and \autoref{eq:minSFR}.
For example, 
the model implies (a) the formation of a $\sim$ few $10^{10}$ \Msun, 
which is larger than the total stellar mass simulated by \citet{lochhaas18}, and (b) a SFR in ORC1C of $\sim$ 300 \Msun/year within the last Gyr.
The total mass is broadly consistent with the masses implied by the modelling in Section \ref{sec:SFR}, but the age of the starburst is somewhat less than the age (a few Gyr) implied by the modelling. On the other hand, the models are poorly constrained by the available photometry. It will be important to obtain better photometric data in the future to improve the models.

A final open question for the star-formation-driven ORC scenario 
concerns the amplitude of the magnetic field.
As discussed above, 
ORC1C is located in a galactic overdensity, but not a rich cluster, suggesting an ambient magnetic field lower than the $8 \ \mu$G
required for CMB equipartition.
However, processes such as adiabatic compression or a local dynamo can amplify the ambient magnetic field in the shock downstream.
We can determine an upper limit to the expected field amplitude by considering energy density equipartition with the thermal energy content of the  material downstream of the termination shock \citep[e.g.,][]{bustard17}, $u_B \leq u_{\rm down,therm}$.
At the position of the shock, the ram pressure of the upstream gas matches the the thermal pressure of the downstream gas,
$p_{\rm up,ram} = p_{\rm down, therm} = 2/3 u_{\rm down, therm}$.
Therefore, the dynamically amplified field satisfies
\begin{equation}
    B_{\rm shock} \leq 0.9 \ \mu {\rm G} \ \sqrt{\frac{\dot{M}_{30} \ v_{1000}}{r^2_{260} \ \Omega_{4 \pi}}}
\end{equation}
where the wind mass flux is normalized 
$\dot{M}_{30} \equiv \dot{M}_{\rm wind}/(30 \ M_\odot $/yr), $v_{1000} \equiv v_{\rm wind}/(\rm 1000 \ km/s)$, $r_{260} \equiv r/$260 kpc, and $\Omega_{4 \pi} \equiv \Omega/{4 \pi}$;
which is consistent with the regime of the low magnetic field solution from 
\autoref{fig:emitThickness} but 
again requires an increase of the ORC1C SFR in the last few 100 Myr.

\section{Conclusion}
We have imaged the first ORC with the Meerkat telescope, resulting in a high resolution, high sensitivity image which shows internal structure that was not apparent in previous imaging. We have also obtained polarisation and spectral index information which provides significant constraints on potential models of ORCs. We have also studied the properties and environment of the central host galaxy. 

The  ORC is consistent with a spherical shell of synchrotron emission about 0.5 Mpc diameter, surrounding an elliptical host galaxy which probably contains both an AGN and significant starburst activity in the past.

We have considered three models that may explain the ORC phenomenon: (a) a spherical synchrotron shell caused by a spherical shock from a cataclysmic event (such as a merger of two supermassive black holes) in the host galaxy, (b) a spherical synchrotron shell caused by a starburst termination shock from past starburst activity in the host galaxy, or (c) radio jets viewed end-on.

Our data cannot easily distinguish between these models. However, we have modelled the starburst termination shock in some detail, and obtain a model which is broadly consistent with the observations.

 None of these models adequately explains the internal radio structure. However, the ORC is located in a group of galaxies about nine times as dense as the ambient space density of galaxies in that region, and some of the galaxies are physically located within the ORC shell. We speculate that the internal structure within the ORC may result from interactions of these galaxies with the starburst wind. 

\section*{Data Availability Statement}
The data underlying this article were accessed from the Meerkat Radio Telescope (proposal number DDT-20200519-RN-01). The derived data generated in this research will be shared on reasonable request to the corresponding author.  FITS files of Figure 1 are available in the online Supplementary material.

\section*{Acknowledgements}

We thank the staff of the Meerkat telescope for their help with obtaining and processing the data used in this paper. The MeerKAT telescope is operated by the South African Radio Astronomy Observatory, which is a facility of the National Research Foundation, an agency of the Department of Science and Innovation.
This work makes use of data products from the Wide-field Infrared Survey Explorer, which is a joint project of the University of California, Los Angeles, and the Jet Propulsion Laboratory/California Institute of Technology, funded by the National Aeronautics and Space Administration.
It also uses public archival data from the Dark Energy Survey (DES) and we acknowledge the institutions listed on \url{https://www.darkenergysurvey.org/the-des-project/data-access/}
This research has made use of the ``Aladin sky atlas'' developed at CDS, Strasbourg Observatory, France \citep{aladin}. It has also made extensive use of TOPCAT \citep{topcat} and Ned Wright's cosmology calculator \citep{nedwright}.
Partial support for LR comes from US National Science Foundation Grant AST 17-14205 to the University of Minnesota.
RMC thanks Cassi Lochhaas, Todd Thompson, and Mark Krumholz  for enlightening conversations and correspondenceand acknowledges 
support from the Australian Government through the Australian Research Council, award
DP190101258 (shared with Prof. Mark Krumholz).

%%%%%%%%%%%%%%%%%%%%%%%%%%%%%%%%%%%%%%%%%%%%%%%%%%



\bibliographystyle{mnras}
\bibliography{main} % if your bibtex file is called example.bib

%%%%%%%%%%%%%%%%%%%%%%%%%%%%%%%%%%%%%%%%%%%%%%%%%%

%%%%%%%%%%%%%%%%% APPENDICES %%%%%%%%%%%%%%%%%%%%%

% Don't change these lines
\bsp	% typesetting comment
\label{lastpage}
\end{document}